\documentclass[lettersize,journal]{IEEEtran}
\usepackage{amsmath,amsfonts}
\usepackage{algorithmic}
\usepackage{pifont} % 提供 \ding 命令
\usepackage{algorithm}
\usepackage{color, xcolor}
\usepackage{array}
\usepackage[caption=false,font=normalsize,labelfont=rm,textfont=rm]{subfig}
\usepackage{textcomp}
\usepackage{stfloats}
\usepackage{url}
\usepackage{verbatim}
\usepackage{graphicx}
\usepackage{makecell} 
\usepackage{cite}
\usepackage{caption}
\usepackage{booktabs}
\usepackage{multirow}
\usepackage{bbding}
\usepackage{amssymb}
\usepackage{graphicx}
\usepackage{pifont}
\usepackage{graphicx}
\usepackage{float}

\usepackage[colorlinks=true, linkcolor=red, citecolor=red, urlcolor=red]{hyperref}
\definecolor{mydarkgreen}{RGB}{0, 128, 0}
\usepackage{subcaption}  % To use subfigures
\usepackage{caption}     % To customize caption styles
\usepackage{graphicx}
\usepackage{caption}
\usepackage{subcaption}
\usepackage{float}
\usepackage{titlesec}

\titlespacing{\section}{0pt}{1.5ex plus 0.5ex minus 0.2ex}{0.8ex}
\titlespacing{\subsection}{0pt}{1.2ex plus 0.3ex minus 0.2ex}{0.5ex}
\begin{document}
\title{{Mitigation of Active Power Oscillation in Multi-VSG Grids: An Impedance-Based Perspective}}
\author{Junjie Xiao,~\IEEEmembership{Graduate Student Member,~IEEE,} Lu Wang,~\IEEEmembership{Member,~IEEE,}\\
Xiong Du,~\IEEEmembership{Member,~IEEE}, Pedro Rodriguez,~\IEEEmembership{Fellow,~IEEE}, and Zian Qin,~\IEEEmembership{Senior Member,~IEEE}
% <-this % stops a space
%\thanks{This paper was produced by the IEEE Publication Technology Group. They are in Piscataway, NJ.}% <-this % stops a space
%\thanks{Manuscript received April 19, 2021; revised August 16, 2021.}
}
%\markboth{IEEE Transactions on Power Electronics}%

\maketitle
\begin{abstract}
Active power oscillations (APOs) frequently arise in inverter-dominated power systems with multiple converters operating under Virtual Synchronous Generator (VSG) control, posing risks to system stability and protection coordination. While various mitigation strategies have been proposed, many rely on prior knowledge of system parameters, offer limited damping performance, or involve complex models that lack physical interpretability—making them difficult to apply in practice. To address these challenges, this paper first introduces a physically intuitive RLC equivalent circuit model to explain the root causes of APOs in both stand-alone (SA) and grid-connected (GC) modes. By mapping inertia, damping, and feeder impedance to capacitive, resistive, and inductive elements, respectively, the model reveals how mismatches among converters lead to inter-unit oscillations characterized by LC resonance. Building on this insight, we propose two mode-specific mitigation strategies: (i) in SA mode, a graph-theory based impedance control ensures proportional reactive power sharing and effectively suppresses APOs; and (ii) {in GC mode, adaptive inertia and damping control with feedforward filtering is designed to reshape transient power dynamics while preserving frequency stability}. The proposed methods are validated through extensive simulations and real-time hardware-in-the-loop experiments, demonstrating their effectiveness in suppressing oscillations and enhancing the robustness of multi-converter power systems.
\end{abstract}


\section{Introduction}
{\IEEEPARstart{D}{istributed} generation (DG) has attracted increasing interest due to its role in modern power systems. In particular, shifting from traditional droop-based control to VSG strategies improves key frequency metrics—such as the rate of change of frequency (RoCoF)—which contributes positively to system stability \cite{xiao2020inertial}. However, this shift introduces oscillatory behavior that complicates system dynamics and may yield pronounced active power oscillations. Such oscillations can appear in both GC and SA modes \cite{5456209}. In severe cases, large transient currents may trigger overcurrent protection, undermining system stability \cite{chen2023active}. A wide range of VSG variants has been developed to alleviate oscillations. Broadly, existing approaches can be grouped into: (i) \emph{model-free} schemes that adapt parameters using measured frequency/power deviations, and (ii) \emph{model-based} enhancements that shape dynamics using structural design or feedforward compensation.

Model-free schemes include adaptive parameter methods, such as \cite{8741094,8798678,7727967,li2023communication}. They typically increase inertia when frequency deviates from nominal and decrease it as synchronization is approached; they can improve transients but introduce nonlinearities and may unintentionally alter pre-designed inertia. In parallel, feedback-based enhancements have been proposed to augment the basic VSG structure. These methods rely on real-time measurements of frequency, power, or phase variations to adjust the control response. For instance, \cite{9321738,10004198,9896833} incorporate transient power deviations into feedback loops to enhance disturbance compensation. Additionally, graph-based secondary control schemes have been developed to achieve frequency consensus under shared authority, with communication constraints explicitly considered \cite{9778180,9220837,10104161}. More recent work, including \cite{gao2024adaptive,10130744}, introduces mutual damping terms based on frequency differences among neighboring DGs.

{Despite their practical benefits, model-free strategies are inherently passive, meaning that they rely on the detection of oscillations before taking corrective action. As a result, VSG systems inevitably experience power and frequency oscillations, particularly when the power control loop is slow. Moreover, the effectiveness of these methods depends heavily on the accuracy of the oscillation detection process.}

In addition, model-based approaches are employed to mitigate APO. For instance, \cite{8419291,7797447} adjusts the damping and inertia coefficients simultaneously in real time to determine and maintain the optimal damping ratio, thereby suppressing power and frequency oscillations throughout the operation. However, this approach changes the preset inertial response of the VSG. By considering both frequency and power dynamics in advance, \cite{8454492} and \cite{8972379} adopt an offline pole–zero design and fix the controller parameters in advance, like the Power System Stabilizer. This method improves the transient performance but limits adaptability under changing operating conditions.

{In \cite{9716746}, the active power reference is feedforwarded to compensate for VSG frequency variations. Similarly, a phase angle feedforward path has been proposed, as in \cite{9712349}. However, in SA mode, feedforward compensation tends to be highly sensitive to noise. In large-scale power systems, where complex interactions among power electronic devices occur, the lack of coordination with this method makes it difficult to effectively suppress oscillations.} {Moreover, \cite{8790737} analyzes the power oscillation mechanism and uses virtual impedance to suppress power oscillations caused by line mismatches. In \cite{10158393}, parameter design principles are defined to eliminate all transient circulating power theoretically. Although these model-based methods provide effective damping, they typically require foreknowledge of the system, such as feeder impedance. This cannot be assured in practical scenarios. {Additionally, considering that converters are often installed far from the bus, variations in line impedance can render carefully tuned parameters ineffective, further compromising the effectiveness.} Once line parameters are uncertain or vary during operation, fixed virtual impedance may lose effectiveness and even induce new oscillations or voltage drops.
\begin{table*}[!t]
\caption{{Comparison of the methods for power oscillation mitigation.}}
\label{Comparison_of_the_methods}
\setlength{\tabcolsep}{5pt} % 调整列间距
\begin{center}
\begin{tabular}{@{}ccccccccccccc@{}}
\toprule
\makecell{Type}                         & Method            & Reference                                                                               &Tuning              & \makecell{SA Mode}                                    & \makecell{GC Mode}          & \makecell{Comms\\ Burden}  & \makecell{Detect\\ Burden}            &\makecell{Fast\\ Response}  &\makecell{Feeder\\ Info free}      \\ \midrule
\multirow{5}{*}{{Model free}} & \makecell{Adaptive Parameter}  &\makecell{\cite{8741094,8798678,7727967,li2023communication} }         &\makecell{Inertia, Damping}       & \makecell{\ding{72}\ding{72}\ding{72}\ding{73}\ding{73}}          &\makecell{\ding{72}\ding{72}\ding{72}\ding{73}\ding{73}}         &\makecell{N/A} & High &\makecell{\scalebox{0.8}{\XSolidBrush}}  &\makecell{\scalebox{1.2}{\checkmark}} & \\
                            \cmidrule(lr){2-11}   
                            & \makecell{Feedback Loop}  &\makecell{\cite{9321738,10004198,9896833}}  &\makecell{Power ref, Freq} & \makecell{\ding{72}\ding{72}\ding{72}\ding{73}\ding{73}}          &\makecell{N/A}                    &\makecell{N/A} &High &\makecell{\scalebox{0.8}{\XSolidBrush}} &\makecell{\scalebox{1.2}{\checkmark}}    \\      \cmidrule(lr){2-11}  
                        & \makecell{Consensus Algorithm}  &\makecell{\cite{9778180,10104161,gao2024adaptive,9220837,10130744}} &\makecell{Power ref, Impedance}             & \makecell{\ding{72}\ding{72}\ding{72}\ding{72}\ding{73}}          &\makecell{N/A}         &\makecell{High} &\makecell{High} &\makecell{\scalebox{0.8}{\XSolidBrush}}&\makecell{\scalebox{1.2}{\checkmark}}    \\         \cmidrule(lr){2-11}  
                            & \makecell{\textcolor{mydarkgreen}{Proposed Method}}  &\makecell{\textcolor{mydarkgreen}N/A}  &\textcolor{mydarkgreen}{\makecell{Imped, Inertia, damping}}   & \makecell{\textcolor{mydarkgreen}{\ding{72}\ding{72}\ding{72}\ding{72}\ding{72}}}          &\makecell{\textcolor{mydarkgreen}{\ding{72}\ding{72}\ding{72}\ding{72}\ding{73}}}                   &\makecell{\textcolor{mydarkgreen}{Low}} &\makecell{Low}  &\makecell{\scalebox{1.2}{\checkmark}}&\makecell{\scalebox{1.2}{\checkmark}} \\               \midrule
\multirow{6}{*}{{Model based}} &\makecell{Optimal Damping}  &\makecell{\cite{8419291,7797447}}  &\makecell{Inertia, Damping}  & \makecell{\ding{72}\ding{72}\ding{72}\ding{73}\ding{73}}          &\makecell{\ding{72}\ding{72}\ding{72}\ding{73}\ding{73}}                                       
 &\makecell{N/A} &\makecell{High} &\makecell{\scalebox{0.8}{\XSolidBrush}}&\makecell{\scalebox{0.8}{\XSolidBrush}}  \\ 
                            \cmidrule(lr){2-11}   
                            &\makecell{Pole–Zero Design} &\makecell{\cite{8454492, 8972379}}  &\makecell{Pole, Zero} 
         &\makecell{\ding{72}\ding{72}\ding{72}\ding{73}\ding{73}}  &\makecell{\ding{72}\ding{72}\ding{72}\ding{72}\ding{73}} &\makecell{N/A} &\makecell{Low} 
                            &\makecell{\scalebox{1.2}{\checkmark}} &\makecell{\scalebox{0.8}{\XSolidBrush}} 
                           \\ \cmidrule(lr){2-11} 
                            &\makecell{Feed Forward} &\makecell{\cite{9716746,9712349}}  &\makecell{Freq, Phase} & \makecell{\ding{72}\ding{72}\ding{72}\ding{73}\ding{73}}          &\makecell{\ding{72}\ding{72}\ding{72}\ding{72}\ding{73}} &\makecell{N/A} &\makecell{Low} &\makecell{\scalebox{1.2}{\checkmark}} &\makecell{\scalebox{0.8}{\XSolidBrush}} \\  \cmidrule(lr){2-11}  
 %%%%%%
  & \makecell{Virtual Impedance} &\makecell{\cite{8790737,10158393}}  &\makecell{Impedance} & \makecell{\ding{72}\ding{72}\ding{72}\ding{73}\ding{73}}          
  &\makecell{N/A}
  &\makecell{N/A} & \makecell{High} &\makecell{\scalebox{1.2}{\checkmark}} &\makecell{\scalebox{0.8}{\XSolidBrush}} \\  \bottomrule
\end{tabular} 
\end{center}
\textbf{Note:} \ding{72} = full star, representing a positive performance (score achieved); \ding{73} = empty star, representing no score in this criterion; {N/A} = not discussed in the referenced method, hence not applicable. 
\end{table*}

Beyond control tuning, a modeling gap persists: state-space models are precise yet opaque for root-cause diagnosis, while equivalent-admittance views aid system-level insight but still obscure how specific control parameters excite oscillations. Equivalent-admittance models \cite{9632345,9970405,9354911} embed inertia/damping into synthetic admittances and provide system-level views, yet they do not isolate how individual control parameters excite oscillations or directly suggest circuit-intuitive suppression rules.

To bridge these gaps, we map VSG dynamics to an RLC network, which directly links control parameters to resonance mechanisms; this model underpins the SA-DVI and GC-adaptive-inertia strategies summarized below. In GC, oscillations are interpreted as second-order RLC resonance; we therefore introduce an adaptive-inertia feedforward that reshapes transients at set-point changes without degrading RoCoF during GC-to-SA transitions. In SA, oscillations stem mainly from parameter mismatches across units; we propose a distributed virtual-impedance (DVI) scheme that aligns system-level impedance ratios, eliminating APOs while preserving nominal inertia/damping. A comparison with recent works is summarized in Table~\ref{Comparison_of_the_methods}.

This paper's main contributions are summarized as follows:

1) The equivalent circuit model of VSG control is proposed, which provides clear physical interpretations. In this model, inertia, damping, and feeder impedance are analogized to capacitance, resistance, and inductance, respectively. Active power reference changes and load switches are excitation sources that inject power into the circuit. The power oscillations are viewed as LC resonance phenomena.

2) A distributed virtual impedance method is proposed to attenuate oscillations in the SA mode. It provides faster response to load variations and reduced communication dependence. This is because the consensus law aligns with equivalent impedances during steady operation, so transient events do not require high-rate message exchanges.

3) Conventional VSG control is modified by adding a feedforward loop that enables adaptive inertia and damping, which suppress APOs while maintaining PCC-frequency quality in GC operation under both strong and weak grid conditions. The adaptation is driven by a high-pass–filtered power-reference signal; thus, it vanishes at steady state and does not affect SA mode operation.}}
\begin{figure}[!htbp]
\centering
\includegraphics[width=0.4\textwidth]{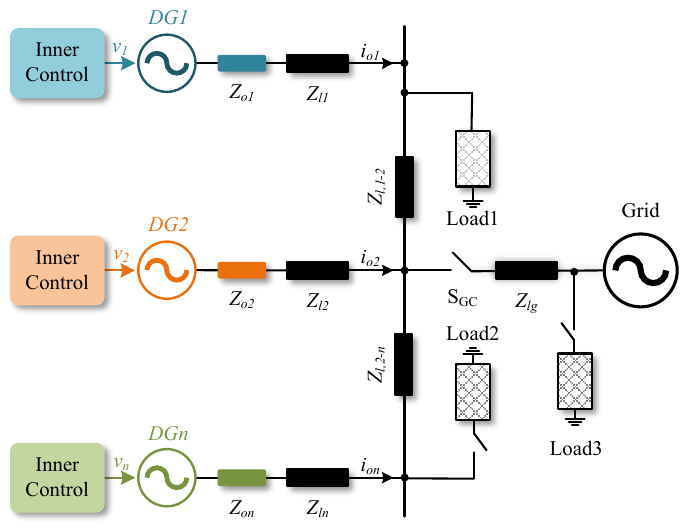}
\caption{The microgrid configuration consisting of $n$ DGs.}
\label{MG}
\end{figure}
\section{{Revisit VSG Control}}
{This section reviews two widely adopted primary control strategies in $n$-inverter-based microgrids, as illustrated in Fig.~\ref{MG}. {Each inverter is operated as a {grid-forming} (GFM) converter: it synthesizes its own voltage angle and magnitude without a PLL.} The inverter is modeled as an ideal voltage source in series with an output impedance \( Z_{oi} \) and a feeder impedance \( Z_{li} \). The impedance between $DG_i$ and $DG_j$ across the bus is denoted by \( Z_{l,i-j} \), and \( Z_{l,g} \) represents the grid impedance. The following standard assumptions are made to simplify the analysis: (i) The model \emph{retains} the dominant outer-loop dynamics, but the inner control loop operates much faster than the outer power loop, so its dynamics are neglected. (ii) The line impedance is predominantly inductive. This assumption is widely valid in medium- and high-voltage systems due to transformer leakage inductance and the length of feeder cables. Virtual impedance is also added to enhance the effective inductive behavior.(iii) The bus impedance \( Z_{l,i-j} \) is considered negligible. This is justified by the fact that the combined feeder and virtual impedances are significantly larger than the low bus impedance, which is typically designed for efficient power transfer. Therefore, variations in bus impedance or load location have limited impact on power sharing. The use of virtual impedance further reinforces this condition by ensuring the dominance of feeder impedance.}
\subsection{Review on Traditional Droop and VSG Control}
{The power control loops of droop and VSG control are shown in Fig.{\ref{droop and VSG}} where the main differences lie in the active power control loops, and the reactive power loops are ignored.
\begin{figure}[H]
\centering
\includegraphics[width=2.5in]{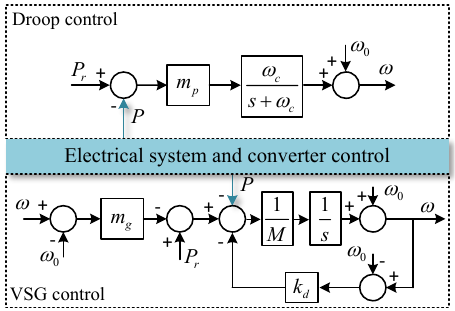}
\caption{Active power loop: droop control and VSG control.}
\label{droop and VSG}
\end{figure} 

The active power loop of droop control is shown as:
\begin{equation} 
\omega = {\omega _0} - {m_p} \frac{{{\omega _c}}}{{s + {\omega _c}}} (P - {P_r})
\label{1}
\end{equation}
where $\omega$ represents the generated angular frequency reference of the inverter output voltage, $\omega_0$ is the nominal value of angular frequency, $m_p$ is the droop coefficient, $P$ represents the inverter output active power, $P_r$ is the nominal value of active power, $m_p$ is the droop coefficient, and $\omega_c$ is the cutoff angular frequency of the low-pass filter.

The active power control equation for VSG is shown in (\ref{2}).
\begin{equation}
{{P_r}-m_\textit{g}(\omega-\omega_0) - P - k_d(\omega - {\omega _0})= M \frac{{d(\omega - {\omega _0})}}{{dt}}} 
\label{2}
\end{equation}
where $m_\textit{g}$ is the proportional coefficient of the governor, $k_d$ is the damping factor, and $M$ is the moment of inertia.

Accordingly, the small-signal model of droop control and VSG can be simplified to Fig.{\ref{Small-signal_droop}} and Fig.{\ref{Small-signal_VSG}}, respectively. The linearized model predictions agree well with nonlinear simulations (Fig.~\ref{Tra_VSG}–Fig.~\ref{Active_power_comparison_GC}), validating its accuracy in capturing the APO dynamics.
\begin{figure}[htbp]
\centering
\includegraphics[width=3in]{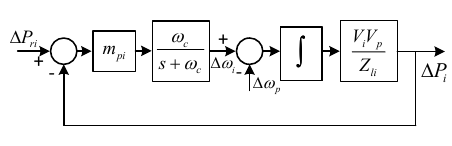}
\caption{Small-signal model of droop control.}
\label{Small-signal_droop}
\end{figure}

In Fig.{\ref{Small-signal_droop}}, $\omega_p$ is the angular frequency of the point of common coupling (PCC). $V_i$ represents the unit output voltage, $V_p$ is the PCC voltage, and $Z_{li}$ denotes the feeder impedance.

Considering the RoCoF requirements, the frequency dynamics when the load changes are the main focus in the SA mode. For droop control, the transfer function of angular frequency change $\Delta\omega$ over loading transition $\Delta P$ is shown in (\ref{3}).
\begin{equation}
G_\textit{{d,sa}}=\frac{\Delta\omega_i}{\Delta P_i}=-m_{pi}\frac{\omega_c}{s+\omega_c}=-m_{pi}\frac{1}{s/\omega_c+1}  
\label{3}
\end{equation}

In the GC mode, the converters are expected to track the power commands precisely. Therefore, the small-signal transfer function from active power reference $\Delta P_{ri}$ to the actual output active power $\Delta P_i$ is considered as shown in (\ref{4}), and $\Delta \omega_p$ is the PCC frequency disturbance. Here, $K_i=V_iV_p/Z_{li}$.
\begin{equation}
\begin{aligned}
  \Delta P=\frac{{{\omega _c}{m_p}K}}{{s^2 + {\omega _c}s+{\omega _c}{m_p}K}} \Delta P_r 
  - \frac{{(s + {\omega _c})K}}{{s^2 + {\omega _c}s + {\omega _c}{m_p}K}} \Delta \omega_p
\end{aligned}
\label{4}
\end{equation}
 
For traditional VSG control in Fig.{\ref{droop and VSG}}, by simplifying the inertial and damping term in Fig.{\ref{droop and VSG}} with $J=M$ and $D=k_d+m_g$. The small signal model of $ith$ VSG {\cite{8454492}} in Fig.{\ref{droop and VSG}} can be simplified as shown in Fig.{\ref{Small-signal_VSG}}.
\begin{figure}[H]
\centering
\includegraphics[width=3in]{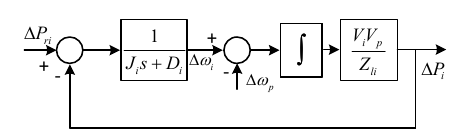}
\caption{Small-signal model of VSG control.}
\label{Small-signal_VSG}
\end{figure}

The VSG small-signal transfer function of angular frequency change $\Delta\omega_i$ { over loads change} $\Delta P_i$ is shown in (\ref{VSG_SA}).
\begin{equation}
G_\textit{{v,sa}}=\frac{\Delta\omega_i}{\Delta P_i}=-\frac{1}{J_is+D_i}=-\frac{1}{D_i}\frac{1}{J_is/D_i+1}  
\label{VSG_SA}
\end{equation}

In GC mode, DGs are expected to track the power commands precisely. Therefore, the transfer function from power reference $\Delta P_{ri}$ to the actual outputs $\Delta P_i$ is considered as: 
\begin{equation}
\Delta {P_{i}} = \frac{{{K_i}}}{{{J_i}{s^2} + {D_i}s+{K_i}}}\Delta {P_{ri}}- \frac{{{K_i}\left( {{J_i}s + {D_i}} \right)}}{{{J_i}{s^2} + {D_i}s + {K_i}}}\Delta {\omega _{p}} 
\label{VSG_GC}
\end{equation}
Combining (\ref{3})-(\ref{VSG_GC}), with $m_p=1/D$, $\omega_c=D/J$, the droop control can be equivalent to VSG control. where $J_i$ and $D_i$ are the inertia and damping factors. In this paper, the VSG is adopted for the power converter control verification.}
\subsection{Active Power Oscillation with VSG control}
As VSG simulates the synchronous generator's inertia and damping characteristics, the oscillation characteristics are inevitably introduced. This subsection investigates the mechanism of active power oscillation in SA and GC modes.
\subsubsection{Oscillation in SA mode}
The active power across the feeder can be obtained as shown in (\ref{7}).
\begin{equation} 
\Delta P_i=\frac{V_iV_p}{Z_{li}}sin \delta_i=K_i\frac{\Delta\omega_i-\Delta\omega_p}{s}
\label{7}
\end{equation}
where $\delta_i$ is the phase difference between $i$-th DG and PCC. Based on the small signal model of VSG, the transfer function from the PCC voltage fluctuates to the $i$-th VSG output  variation characteristics can be expressed as (\ref{8}):
\begin{equation}
\left\{
\begin{aligned}
\Delta P_{i} &=\frac{\displaystyle{-K_{i}(J_is + D_i)}}{\displaystyle{J_is^2 + D_is + K_{i}}}\Delta\omega_p\\
\Delta \omega_{i} &= \frac{\displaystyle K_{i}}{\displaystyle{J_is^2 + D_is + K_{i}}}\Delta\omega_p\\
\end{aligned}
\right.
\label{8}
\end{equation}

The output power of the involved DGs is equal to the load power $\Delta P_L$, which can be represented as in (\ref{9}).
\begin{equation}
\label{9}
\sum\limits_{i = 1}^{n} \Delta P_i=\Delta P_L  
\end{equation}
{where $n$ is the number of the involved converters.} By combining equations (\ref{8})-(\ref{9}), the transfer function that describes the interaction between load changes and PCC frequency variations for VSGs under SA mode can be derived as shown in (\ref{10}). This derivation facilitates calculating the PCC frequency responses of VSGs under varying load conditions.
\begin{equation}
\label{10}
\Delta\omega_p=-\frac{1}{\sum\limits_{i = 1}^{n}G_i(s)(J_is+D_i)}\Delta P_{L} 
\end{equation}
where $G_{i}= {K_{i}}/(J_is^2 + D_is + K_{i})$. For a multi-VSG scenario, the load change causes a variation in PCC frequency. As shown in (\ref{8}), the different transfer functions from PCC frequency to DG's outputs lead to a DG's dynamic disparity, which contributes to the oscillations.
By combining equations (\ref{8})-(\ref{10}), the transfer function describes the dynamics and steady state of the DGs when the load switch is shown as in (\ref{11}).
\begin{equation}
\left\{
\begin{aligned}
\Delta \omega_i&=-\frac{\displaystyle G_i(s)}{\sum\limits_{k = 1}^{n}G_k(s)(J_ks+D_k)}\Delta P_{L}\\
\Delta P_i&=\frac{\displaystyle G_i(s)(J_is+D_i)}{\sum\limits_{k = 1}^{n}G_k(s)(J_ks+D_k)}\Delta P_{L}\\
\end{aligned}
\label{11}
\right.
\end{equation}

{Accordingly, increasing $D$ or decreasing $J$ improves the damping ratio, thereby mitigating oscillations during load changes in multi-VSG systems.} However, since $D$ is coupled with the droop coefficient, representing the steady-state frequency deviation, modifying $D$ will inevitably change the frequency deviation nadir. Additionally, decreasing $J$ is undesirable for VSGs as it may violate the RoCoF rules. Consequently, a trade-off between the dynamic and steady-state performance of the VSG is unavoidable. Moreover, existing virtual impedance techniques may be ineffective, as exact parameter matching of parallel VSGs may not be entirely achievable. While inserting substantial virtual impedance into the control loop can mitigate oscillations, it also leads to considerable and unexpected voltage drops.
\subsubsection{Oscillation in GC mode}
In the grid-connected mode, the active power is determined by the power reference $\Delta {P_{ri}}$ and PCC frequency $\Delta {\omega _{p}}$, which is shown in (\ref{VSG_GC}).

Based on (\ref{7}), the power support from the utility grid is (\ref{13}).
\begin{equation}
\Delta {P_\textit{g}} =-\frac{{{K_\textit{g}}}}{s}\Delta {\omega _{p}}
\label{13}
\end{equation}
Where ${P_\textit{g}}$ is the utility grid power output, and ${K_\textit{g}}=V_gV_p/Z_{lg}$. Assuming the load remains unchanged, the sum of power flow change is zero, denoted as (\ref{14}).
\begin{equation}
\label{14}
\sum\limits_{i = 1}^{n} \Delta P_i+\Delta {P_\textit{g}}=0              
\end{equation}

{{Based on (\ref{VSG_GC}), (\ref{13})-(\ref{14}), the transfer function from power reference to PCC frequency is as (\ref{15}).}
\begin{equation}
\label{15}
\Delta\omega_p=\frac{\sum\limits_{i = 1}^{n}G_i(s)\Delta P_{ri}}{K_\textit{g}/s+\sum\limits_{i = 1}^{n}G_i(s)(J_is+D_i)}  
\end{equation}

Accordingly, the frequency and active power dynamics when power commands change is simplified as in (\ref{16})
\begin{equation}
\left\{
\begin{aligned}
\Delta \omega_i&= sG_i(s)\Delta {P_{ri}}/K_i +G_i(s)\Delta{\omega _{p}}\\
\Delta P_i&=G_i(s)\Delta P_{ri}-G_i(s)(J_is+D_i)\Delta \omega_{p}\\
\end{aligned}
\label{16}
\right.
\end{equation}
}
Based on (\ref{15}) and (\ref{16}), when the VSG operates in GC mode, it is typically modeled as a second-order system \cite{9716746}.
{According to classical control theory, increasing the damping ratio---by reducing \(J\) or increasing \(D\)---helps suppress power oscillations and improves transient response. However, in SA mode, a smaller \(J\) may lead to excessive RoCoF, threatening frequency stability and violating grid codes. Likewise, while a larger \(D\) enhances damping, it narrows the power control bandwidth and slows reference tracking. Moreover, \(1/D\) defines the droop coefficient, so \(D\) must balance between dynamic performance and steady-state frequency deviation. Raising the feeder impedance \(Z_l\) also helps decouple unit dynamics but causes greater voltage drops, especially under high-power conditions.

These trade-offs imply that parameter settings optimal for oscillation suppression in grid-connected mode may compromise frequency stability in islanded operation---motivating a coordinated design across both modes.}
\section{Equivalent Impedance Circuit of VSG}
An equivalent circuit model is developed in this section to learn the root cause of APO issues intuitively, which, in the end, leads to the proposed mitigation measures for APO issues in both SA mode and GC mode.
\subsection{Single-VSG Equivalent}
From the VSG small signal model in Fig.{\ref{Small-signal_VSG}}, the dynamic interaction is as in (\ref{17}) and (\ref{18}):
\begin{equation} 
\frac{1}{{{K}_i}}\frac{{d\Delta P_i}}{{dt}} = \Delta \omega_i - \Delta {\omega}_p  
\label{17}
\end{equation}
\begin{equation} 
\Delta P_{ri}=\Delta P_i+J_i\frac{\Delta\omega_i}{dt}+D_i\Delta\omega_i
\label{18}
\end{equation}

Tab.{\ref{tableI}} shows the analogy relationships between the control and circuit variables. Where subscript $i$ represents the DG$i$'s parameter, while subscript $\textit{g}$ represents those of the utility grid, and subscript $p$ represents the PCC's parameter. 
\begin{table}[htbp]
\setlength{\tabcolsep}{3pt} % 调整列间距
\setlength{\tabcolsep}{3.5pt} % 调整列间距
\renewcommand{\arraystretch}{1.5} 
\begin{center}
\centering
\caption{Correspondence between VSG and circuit variables}
\label{tableI}
\begin{tabular}{ccccccccccc}
\toprule
 & \text{circuit} & ${U_i}$& ${U_p}$ & ${I_i}$& ${I_{ri}}$& ${I_L}$ & ${R_i}$  & ${C_i}$ & ${L_i}$ & ${L_g}$\\ \midrule
 & \text{VSG} & $\Delta\omega_i$ & $\Delta\omega_p$&  $\Delta P_i$ &$\Delta P_{ri}$&  $\Delta P_L$& $1/D_i$&$J_i$&$1/K_i$ &$1/K_g$ \\ 
\bottomrule
\end{tabular}
\end{center}
\end{table}

With the analogy, $\Delta\omega_i$ is equivalent to the voltage $U_i$, representing the frequency change. $\Delta P_i$ is the equivalent to the current $I_i$, representing the active power change. $J_i$ is equivalent to a capacitance $C_i$, is the inertia coefficient; $1/D_i$ is the resistance $R_i$, is the damping factor; $1/K_i$ is equivalent to an inductance $L_i$, representing  feeder impedance term. According to Tab.{\ref{tableI}}, (\ref{17}) and (\ref{18}), the VSG model can be equivalent to the circuit model, as shown in (\ref{19}) and (\ref{20}). 
\begin{equation} L_i\frac{{d{I_i}}}{{dt}} ={U_i}-{U}_p  
\label{19}
\end{equation}
\begin{equation} 
{I}_{ri}=I_i+C_i\frac{dU_i}{dt}+\frac{U_i}{R_i}
\label{20}
\end{equation}

As shown in (\ref{9}), the current should follow $\sum\limits_{i=1}^{n}  I_i= I_L$.

\begin{figure}[htbp]
\captionsetup[subfigure]{font=small} 
\begin{minipage}{0.33\textwidth}
\centering
\subfloat[Equivalent in SA mode]{\includegraphics[width=0.65\textwidth]{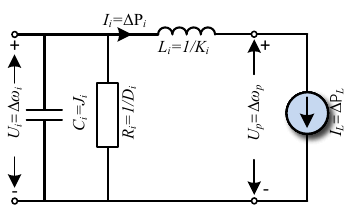}}
\subfloat[Equivalent in GC mode]{\includegraphics[width=0.8\textwidth]{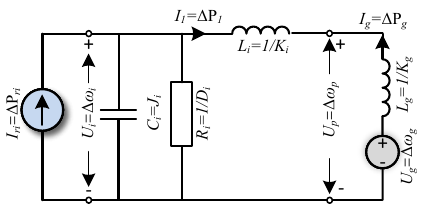}}
\end{minipage}
\caption{Single VSG equivalent circuit.}
\label{VSGequivalent}
\end{figure}

Combining (\ref{8})(\ref{9}) and (\ref{19})(\ref{20}), the VSG models in SA and GC modes can be analogized the equivalent circuit in Fig.\ref{VSGequivalent}, where $I_{ri}$ and $I_L$ can also be viewed as two excitation sources. Accordingly, the power–frequency relationship in the VSG is analogous to the current-voltage relationship in a second-order RLC circuit. The inertia coefficient $J_i$ suppresses frequency changes similarly to how the capacitor stabilizes circuit voltage. The damping coefficient $D_i$ governs the angular frequency changes in output power, analogous to how resistance determines the voltage change in the circuit.

{In the SA mode, the current source $I_L$ is enabled when the load switches. Consequently, the current $I_i$ increases to match $I_L$. The capacitance $C_i$ reduces the voltage change rate $U_i$, analogous to VSG providing inertia and maintaining the RoCoF. The steady-state value of $U_i$ is determined by the resistance $R_i$, which acts as the droop coefficient that dictates the frequency deviation. {As compared in Section II, a key distinction between traditional droop control and VSG control is the inclusion of capacitance in the latter.}

In GC mode, the converter current $I_i$ closely follows its reference $I_{ri}$, so the VSG outputs active power as dictated by the control target. Assuming the power grid behaves ideally (i.e., $U_{\textit{g}} = 0$), an increase in $I_{ri}$ doesn’t instantly settle; instead, it can cause short-lived oscillations in $I_i$ before stabilizing. This transient behavior is a result of $RLC$ resonance, a typical feature of second-order systems. As the system evolves, the voltage across the capacitor $U_i$ adjusts accordingly and trends toward zero, helping the point of common coupling (PCC) frequency align with that of the grid. On the other hand, droop-based control—lacking a capacitive path—generally avoids such oscillations; with no $C_i$, there’s no $RLC$ structure, so the current tends to follow the reference more smoothly.}
\subsection{Multi-VSG Equivalent}
In this section, the transient of multi-VSG operating in SA mode is intuitively analyzed through the resonance in its equivalent circuits. The current expression for each branch can be obtained by establishing nodal voltage equations for circuit analysis. Subsequently, the resonance in the equivalent circuit can be analyzed, providing insights for deriving circuit parameter configuration rules. Similarly, VSG parameters can be configured to eliminate power oscillations during the VSG's transient. Considering the frequency stability requirements, the selection of VSG parameters gains physical significance.
\begin{figure}[H]
\centering
\includegraphics[width=3.4in]{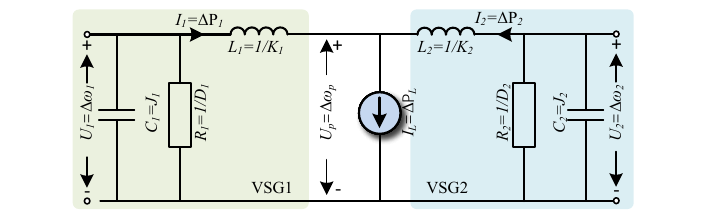}
\caption{Multi-VSG's equivalent circuit perspective in SA mode.}
\label{Multi VSG in SA mode}
\end{figure}

\textbf{\textit{SA Mode:}} Based on Fig.\ref{VSGequivalent}(a), the multi-VSG equivalent circuit is derived. { We illustrate with two converters for clarity; however, equations (\ref{7})-(\ref{16}) are written for general $n$, so all conclusions extend directly to $n$ parallel VSGs.}  They can be equivalent to the circuit as shown in Fig.{\ref{Multi VSG in SA mode}}. The interaction $I_1 + I_2 = I_L$ holds throughout the entire operation. This indicates that the circuit operates in parallel, and the current sharing ratio $I_1$:$I_2$ is determined by the equivalent impedance of each branch, where the impedance $Z_{ei}$ is shown in (\ref{21}). 
\begin{equation} 
{Z_{ei}=\frac{U_p}{I_i}=\frac{1}{{C_is + 1/R_i}} +sL_i}
\label{21}
\end{equation}

{Given a current source, the current \( I_i \) flowing into each branch is determined by its corresponding impedance. Discrepancies in impedance characteristics, represented by resonance mismatches, can result in disproportional current sharing. This sharing behavior can be divided into (1) steady-state, governed by the proportionality of resistive components \( R_i \) aligned with damping coefficients \( D_i \), and (2) dynamic, which requires the equivalent impedances \( Z_{ei} \) to remain proportional over the frequency range of interest.

Importantly, system-level oscillations are not necessarily caused by local resonance within individual converters. As long as the overall impedance ratios are properly coordinated, proportional current sharing can be maintained—even if individual branches exhibit resonance peaks.}

{Accordingly, the circuit elements should be tuned proportionally to avoid oscillation. Converting to the VSG control variable in Tab.{\ref{tableI}}, the VSG parameters can be set as in (\ref{23}). Here, $P_{mi}$ denotes the maximum output active power capacity of $i$-th converter.}
\begin{equation} 
{\frac{P_{mi}}{P_{mj}}=\frac{J_i}{J_j}=\frac{D_i}{D_j}=\frac{K_i}{K_j}=\frac{Z_{lj}}{Z_{li}}}
\label{23}
\end{equation}

{To further validate the theoretical correctness of our analysis, we investigate the frequency-domain characteristics of active power dynamics in a two-VSG system under different configurations. Specifically, we compare two cases using the transfer function defined in Eq.~(\ref{11}), which describes the response from load perturbations to each VSG’s output power.

In the first case, as shown in Fig.\ref{bode_two_VSG}, we set \(J_1=300, J_2=600\), and \(D_1=300, D_2=600\), with non-proportional line impedances \(Z_{l1}=Z_{l2}=4.4\,{mH}\). This configuration produces a resonance peak around \(10\,{rad/s}\), indicating potential oscillatory modes. Power sharing corresponds to the damping coefficient ratio consistent with the expected droop behavior in the low-frequency steady-state region.
\begin{figure}[htbp]
\centering
\includegraphics[width=0.7\linewidth]{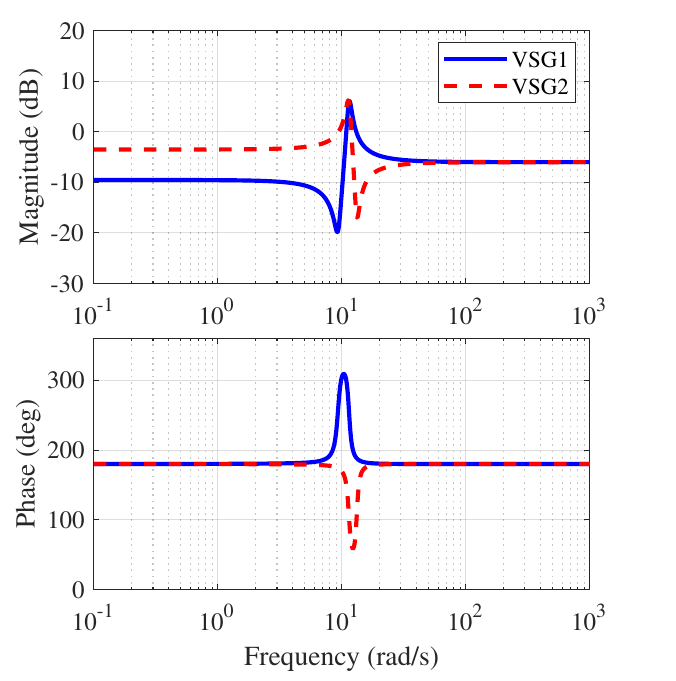}
\caption{{Bode diagrams of the transfer functions ($\Delta P_i$/$\Delta P_L$) for two VSG systems.}}
\label{bode_two_VSG}
\end{figure}

For comparison, we consider a second case where the virtual parameters \( J \) and \( D \) are unchanged. However, the line impedances are proportionally configured, where \( Z_{l1} = 4.4\,{mH} \) and \( Z_{l2} = 2.2\,{mH} \). Under this condition, (\ref{11}) simplifies to:
\begin{equation} 
\Delta P_1 = \frac{P_{m1}}{P_{m1} + P_{m2}} \Delta P_L, \quad 
\Delta P_2 = \frac{P_{m2}}{P_{m1} + P_{m2}} \Delta P_L
\label{proportional parameters power response}
\end{equation}

The resulting transfer functions become frequency-independent constants, and the corresponding Bode plots, omitted for brevity, are flat in magnitude and phase.

These comparisons confirm that when both control parameters and feeder impedances are proportionally tuned, low-frequency active power oscillations are significantly suppressed. This further supports the validity of our proposed strategy from a frequency-domain perspective.}

{Although our main analysis assumes inductive feeders, similar oscillations arise in resistive networks—primarily in reactive power and frequency. We show that proportional active-power sharing still suppresses these oscillations, so the coordination strategy naturally extends to resistive cases.}

\begin{figure}[H]
\centering
\includegraphics[width=3.5in]{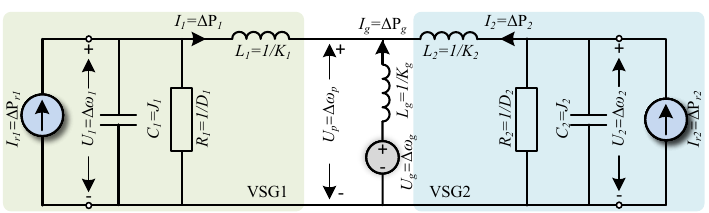}
\caption{Multi-VSG's equivalent circuit perspective in GC mode}
\label{Multi VSG in GC mode}
\end{figure}

\textbf{\textit{GC Mode:}} Based on Fig.\ref{VSGequivalent}(b), two DGs in GC mode can be equivalent to the circuit shown in Fig. \ref{Multi VSG in GC mode}. In this mode, the sum of power support should be the same as the load consumption, denoted as $\sum\limits_{i=1}^{n} I_i+I_\text{g}=0$. Based on Fig.{\ref{Multi VSG in GC mode}}, the power flow across each DG can be intuitively shown. When the current reference step, its influence on current and voltage variation can be shown in (\ref{24}). 
\begin{equation}
\left\{
\begin{aligned}
    U_p&=I_{g} \cdot sL_g\\
    I_{i}&=\frac{1/(L_i+L_\text{g})}{{C_is^2+s/R_i+1/(L_i+L_\text{g})}}\cdot{I_{ri}}
\\
\end{aligned}
\label{24}
\right.
\end{equation}

{When the grid-side inductance \( L_g \) is relatively small, disturbances originating from the reference current \( I_{r1} \) are mostly confined to the local loop formed by \( L_1 \) and \( C_1 \), exerting little effect on VSG2 through \( L_2 \). As a result, the system can be regarded as a loosely coupled second-order structure, where oscillations remain largely isolated within VSG1. As \( L_g \) becomes larger, however, the electrical coupling between VSG1 and VSG2 intensifies. The injected current associated with \( I_{r1} \) begins to spread through both \( L_g \) and \( L_2 \), thereby increasing inter-unit interaction. Under such conditions, the second-order simplification and the assumption of negligible grid feedback become less valid, potentially leading to misinterpretation of system behavior. To address this, our equivalent circuit framework provides a consistent and physically meaningful description under varying \( L_g \) values. A critical factor in both regimes is the high-frequency response of the capacitor \( C_1 \). During transient events, a reduced effective capacitance tends to shift the current path through \( L_1 \), which helps dampen local oscillations.}

An intuitive method is to increase the $L_i$, mitigating the resonance. However, it may cause a sizeable equivalent impedance of the feeder. Another way is to reduce the capacitance $C_i$, which is a possible way to minimize the oscillation; however, the existing associated methods may degrade the  RoCoF in SA mode.
\section{Proposed Control Design}
\subsection{{Power Oscillation Mitigation in SA Mode}}
{In SA mode, active power oscillations in a multi-VSG system can, in theory, be eliminated by proportionally setting the parameters in (\ref{23}). While the inertia and damping coefficients can be configured by keeping the same ratio between \( J \) and \( D \) across all units, it is often difficult to realize the ideal impedance ratio among different VSGs due to uncertainties in actual line inductance. To address this issue, this section suggests mitigating oscillations by tuning the virtual impedance. By doing so, the equivalent output impedance of each unit can be adjusted to reduce dynamic interactions. In the equivalent circuit model shown in Fig.~\ref{Multi VSG in SA mode}, the inductance \( L_i \) is tunable, while the capacitance \( C_i \) and resistance \( R_i \) can be configured according to the DG's output capacity and RoCoF requirements. It is worth noting that mismatched equivalent impedance leads to uneven reactive power sharing. Therefore, proportional reactive power sharing can serve as an indicator of well-aligned equivalent impedance. When the load changes, systems with well-tuned impedance exhibit mitigated active power oscillations.}

As the feeder is assumed to be inductive, the  power flowing through the feeder impedance results in a voltage drop $\Delta V_i$, which can be expressed as:
\begin{equation}
\label{28}
 {\Delta}V_i\approx\frac {X_{ei}Q_i}{V_{i}}
\end{equation}
where $X_{ei}$ is the equivalent impedance of $i$-th VSG. In \cite{10287544,10845014,10547193}, it is demonstrated that the virtual impedance design enables modification of the equivalent feeder impedance $X_{ei}$. This adjustment promotes proportional reactive power sharing.
\subsubsection{Communication network modelling}
{The microgrid's communication network is represented using an undirected cyber graph, illustrating how converters exchange information with neighboring units\cite{10979334}. Each VSG communicates with its adjacent VSGs through the communication network. It can be described by the communication adjacency matrix $A = [a_{ij}]_{n \times n}$. $a_{ij}$ is set to $1$ if units $i$ and $j$ are regularly communicating, and $0$ otherwise. The degree of vertex $\zeta_i$ is defined as $d_i = \sum_{j=1}^n a_{ij}$. The corresponding degree matrix is $D_M = \text{diag}(d_1, \ldots, d_n)$. The Laplacian matrix $L$ of the communication network is defined as $L = D_M - A$.}
\subsubsection{Graph Theory-Based Virtual Impedance implementation}
\begin{figure}[h]
\centering
\includegraphics[width=3in]{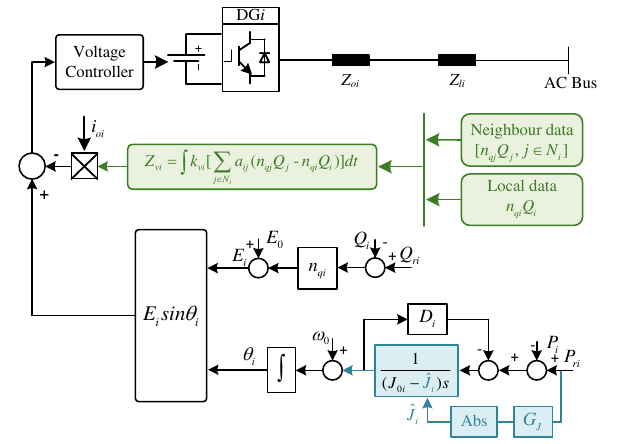}
\caption{Control structure of the proposed method.}
\label{Proposed control}
\end{figure}

The proposed method is shown in Fig.\ref{Proposed control}. The inverters exchange information related to reactive power $(n_{q1}Q_1,\ldots, n_{qN}Q_N)$ with their adjacent units to achieve a reactive power consensus when the impedance has been appropriately adjusted. The reshaped consensus algorithm-based virtual fundamental impedance is expressed as (\ref{30}).
\begin{equation}
\label{30}
{{Z}_{vi}=\int k_{vi}
[\sum\limits_{j\in{N_i}}a_{ij}(n_{qj}Q_j-n_{qi}Q_i)]dt}
\end{equation}
The parameter \( k_{v,i} \) determines the bandwidth of the virtual impedance loop. The adaptively adjusted impedance $Z_{v, i}$ is influenced by neighboring information and the local unit's state. {For the system development or new converter plug-in, reactive power is improperly allocated, and the consensus algorithm prompts the controller to adjust the virtual impedance through the communication network.} This modification aims to achieve balanced reactive power distribution, ensuring proportional sharing across equivalent impedance.
\subsubsection{{Consensus Analysis}}
{{With the graph theory-based virtual impedance, combining (\ref{28}) and (\ref{30}), the reactive power can be expressed as:} 
\begin{equation}
\label{eq_A1}
{{Q_i} = \frac{{{V_i}\Delta {V_i}}}{{\int {{k_{vi}}} \sum\limits_{j \in {N_i}} {{a_{ij}}} ({n_{qj}}{Q_j}-{n_{qi}}{Q_i})dt + {Z_{li}}}}}
\end{equation}
{{where the feeder impedance $Z_{li}$ should be larger than the virtual impedance to ensure stability. If necessary, an initial fixed virtual impedance can be added to guarantee that the total initial impedance is sufficiently large.} Based on (\ref{eq_A1}), the dynamic of the virtual impedance is shown as}
\begin{equation}
\label{eq_A2}
{{\dot Q_i} =  - \frac{{{V_i}\left| {\Delta {V_i}} \right|{k_{vi}}}}{{{{[\int {{k_{vi}}(-L{n_{qi}}{Q_i})dt}  + {Z_{li}}]}^2}}}L{n_{qi}}{Q_i}}
\end{equation}
{To analyze its convergence, we consider the Lyapunov function candidate in (\ref{eq_A3}).}
\begin{equation}
\label{eq_A3}
{\nu(x)=\frac{1}{2}x^{T}Lx}
\end{equation}
{where $x$ is the state $n_{qi}Q_i$ in an simplified expression. Combining (\ref{eq_A2}) and (\ref{eq_A3}), the derivative in time of $\nu(x)$ along the trajectories the controllers is given by:}
\begin{equation}
\begin{split}
{\dot{\nu}(x)}&={x^{T}L\dot{x}}\\
&{= x^TL\frac{{- {n_{qi}}{V_i}\left| {\Delta {V_i}} \right|{k_{vi}}}}{{{{[\int { - {k_{vi}}} Lxdt + {Z_{li}}]}^2}}}Lx}
\end{split}
\label{eq_A4}
\end{equation}
{In eq.(\ref{eq_A4}),  as $L$ is the Positive semi-definite matrix. $\dot{\nu}(x)<0$ when $x_1\neq x_2\neq \cdots \neq x_n$ at the steady state. This implies that the chosen Lyapunov function is globally asymptotically stable. Hence, $\nu(x)$ ultimately tends to zero, making the state error converge to zero and enabling the reactive power to be shared proportionally by the virtual impedance.}}
\subsection{Power Oscillation Mitigation in GC mode}
An excessively large moment of inertia $J$ can reduce the frequency fluctuations; however, it also leads to increased power oscillations. Conversely, selecting a moment of inertia $J$ that is too small degrades frequency stability and inertia response. Similarly, improper selection of the damping coefficient $D$ negatively impacts oscillation and frequency deviation.

In conclusion, a more significant moment of inertia $J$ is preferred in SA mode to maintain the RoCoF, but it increases the oscillation among multi-VSGs. A smaller inertia $J$ is necessary to mitigate oscillations in GC mode for a strong grid, particularly in scenarios involving reference changes. Accordingly, the choice of inertia coefficient can be shown as in (\ref{32}). 
\begin{equation}
\label{32}
{{J_i=J_{0i}-|\frac{\mu s}{\tau s+1} {{\Delta P_{ri}}}|}}
\end{equation}

{In Fig.\ref{Proposed control}, $G_J$ represents a high-pass filter with the transfer function ${\mu s}/({\tau s+1})$. This filter extracts the high-frequency components of the power reference signal, specifically the rapidly changing parts, and feeds them forward to the inertia adjustment link. The parameter selection is shown Table~\ref{Parameter_design_guideline}. This method maintains the RoCoF in the SA mode by allowing inertia to change only when the power reference varies in the GC mode. From the impedance circuit perspective in Fig.\ref{Multi VSG in GC mode}, it can be viewed as changing the capacitance $C_{i}$ in real time under GC mode. Notably, in this context, we assume that $|P_{ri}/\tau| \leq J_{0i}$ for simplicity of expression. If this requirement is not satisfied, a limiter can be introduced to constrain $|P_{ri}|$ within the allowable range.}
\begin{figure*}[b]
\centering
\begin{minipage}[t]{0.32\textwidth}\vspace{0pt}
  \centering
  \includegraphics[height=0.22\textheight]{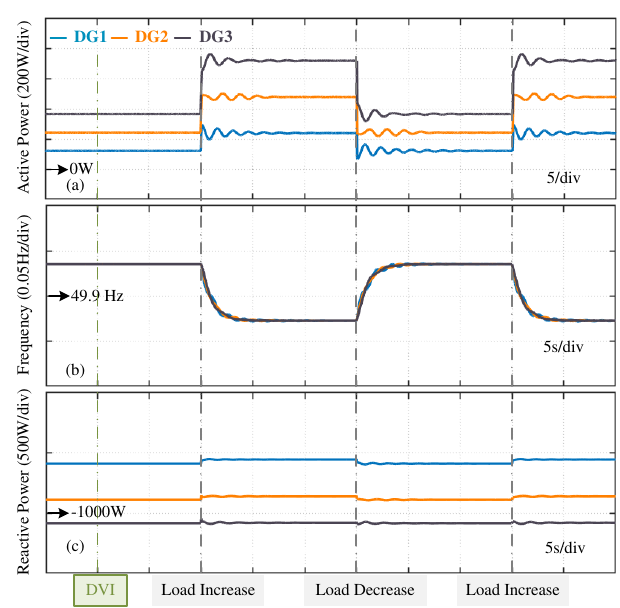}
  \caption{{Dynamics of the traditional VSG.}}
  \label{Tra_VSG}
\end{minipage}
\hfill
\begin{minipage}[t]{0.32\textwidth}\vspace{0pt}
  \centering
  \includegraphics[height=0.22\textheight]{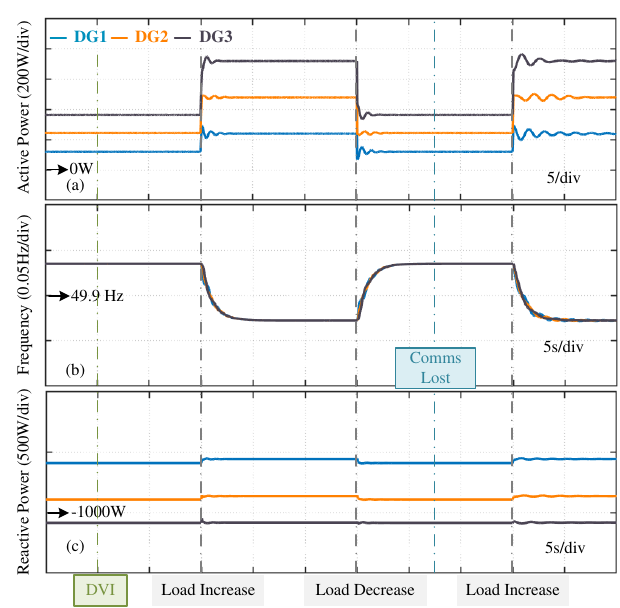}
  \caption{{Dynamics of the DSC in \cite{9220837,gao2024adaptive,10130744}.}}
  \label{DSC}
\end{minipage}
\hfill
\begin{minipage}[t]{0.32\textwidth}\vspace{0pt}
  \centering
  \includegraphics[height=0.22\textheight]{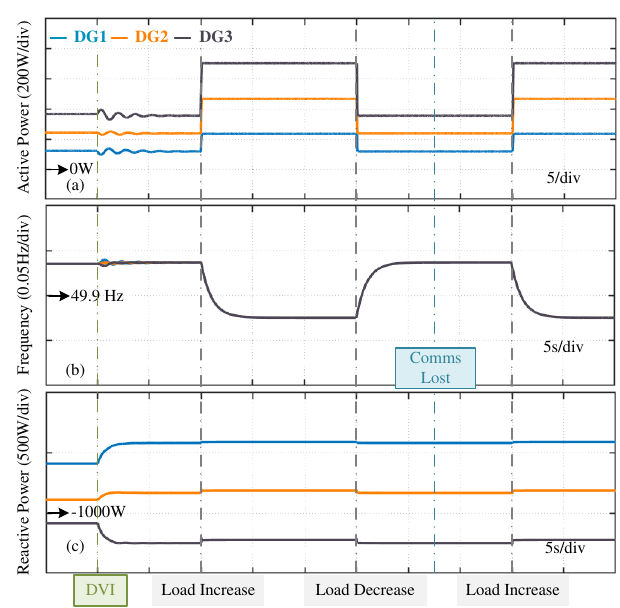}
  \caption{{Dynamics of the proposed DVI.}}
  \label{DVI}
\end{minipage}
\end{figure*}

{Notably, decreasing the converter inertia is acceptable under strong-grid conditions, which are the primary operating scenario considered in this paper. To maintain good performance under weak grids, we introduce a feedforward adaptive damping loop:
$D_i = D_{0i} + \left|\Delta P_{ri}\, {\mu s}/(\tau s + 1)\right|$.
During fast transients, this loop temporarily increases the effective damping, thereby mitigating power oscillations while preserving frequency stability. Because the high-pass filter has zero DC gain, the steady-state damping remains unchanged.}
\begin{table}[htbp]
\caption{{Parameter design guideline.}}
\label{Parameter_design_guideline}
\centering
\setlength{\tabcolsep}{1.8pt} 
\renewcommand{\arraystretch}{1} 
\begin{tabular}{@{}cccc@{}}
\toprule
\makecell{Mode} & Variable  & Symbol & Direct formula  \\ \midrule
\multirow{1}{*}{{SA Mode}} 
& \makecell{\textit{inv i} inertia\\ \textit{inv i} damping\\Consensus gain} 
& \makecell{$J_{0i}$ \\$D_{0i}$\\$k_{v,i}$}  
& \makecell{$J_{0i} = {\Delta P_{\max i}}/{\mathrm{RoCoF}_{\max}}$\\
            $D_{0i} = {\Delta P_{\max i}}/{\Delta \omega_{\max}}$\\
            $k_{v,i} = {\rho\,\omega_{c,\mathrm{VSG}}}/[{n_{q,i}\lambda_2(L)|H(0)|}]$} 
       \\ \midrule
\multirow{1}{*}{{GC Mode}} 
& \makecell{HPF time constant\\HPF gain} 
& \makecell{$\tau$\\ $\mu$}  
& \makecell{$\tau={1}/({k_{\mathrm{HP}}\omega_{c,\mathrm{VSG}}}$)\\
            $\mu/\tau=1,(\mu=\tau)$}  \\ 
\bottomrule
\end{tabular}
\end{table}
\subsection{Parameter Design Guideline of the Proposed Method}
{To facilitate practical implementation, the main parameters of the proposed method are summarized in Table~\ref{Parameter_design_guideline}. The virtual inertia $J_0$ and damping $D_0$ are determined from the maximum power variation $\Delta P_{\max}$ and the desired limits of RoCoF and frequency deviation. This ensures adequate inertial and damping support during transients. In this paper, we set $J = D = 300$, the cutoff frequency $\omega_{c,\mathrm{VSG}}=D_0/J_0=1$. The high-pass filter parameter $\tau$ is selected as $\tau = 1/(k_{\mathrm{HP}}\omega_{c,\mathrm{VSG}})$ with $k_{\mathrm{HP}}\in[10,30]$, so that its cutoff frequency remains well above the VSG loop bandwidth and does not interfere with the fundamental control dynamics. In this paper, we choose $k_{\mathrm{HP}}=10$, and $\mu=\tau=0.1$. In this way, the filter behaves as a standard first-order HPF with unity high-frequency gain. The consensus gain $k_{v,i}$, which determines the bandwidth of the impedance loop, is slower than the VSG bandwidth. It is given by  $k_{v,i}[{n_{q,i}\lambda_2(L)|H(0)|}]$=${\rho\,\omega_{c,\mathrm{VSG}}}$, where $\rho\in[0.01,0.50]$ ensures that this loop operates slower than the reactive-power control loop, thereby avoiding adverse interactions. It set proportional to the VSG bandwidth $\omega_{c,\mathrm{VSG}}$ and scaled by the algebraic connectivity $\lambda_2(L)$ and $n_q$. $|H(0)|=\big|{V_i^0(V_i^0-V_p^0)}/{(Z_{v,i}^0+Z_{l,i}^0)^2}\big|$ represents the magnitude of the small-signal transfer gain from impedance to reactive power.  Herein, we set $\rho=0.05$, and therefore obtain $k_{v,i}=0.01$.}
\section{Verification}
The proposed strategy has been tested in Simulink to validate its effectiveness, where three inverters connected in parallel are considered. In this microgrid system, the output side of the inverters is connected to the AC bus through an LC filter and line impedance. The expected active power-sharing ratio is assumed to be 1:2:3, and {the verification parameters for both simulation and experiments are shown in Tab.\ref{parameter_for_verification}.} {In this paper, a RoCoF threshold of 1 $rad/s^2$ is assumed, as it falls within the typical range of minimum and maximum values reported in \cite{bollen2011integration}. However, this threshold may be adjusted based on practical requirements.}
\begin{table}[htbp]
    \centering
    \small
    \caption{{{Parameters for Verification}}} 
    \label{parameter_for_verification}
    \setlength{\tabcolsep}{2pt} % 调整列间距
    \renewcommand{\arraystretch}{1} % 调整行高
    %\scriptsize % 调整字体大小
    \begin{tabular}{l c c} % 修改为正确的3列格式
        \toprule
        \textbf{Symbol} & \textbf{Description} & \textbf{Value} \\
        \midrule
        $U_{dc}$ & DC source voltage &  300 V \\
        $f_s$ & Switch frequency & 20$k$ Hz \\
        $L_f$  & Inductor of LC filter & 2.2 $m$H \\
        $C_f$  & Capacitor of LC filter  & 12 $\mu$F \\
        $\omega_{0}$  & Nominal angular frequency & 314 rad/s \\
        $V_0$  & Nominal voltage amplitude & 190 V \\
        $k_{vi}$  & Virtual impedance loop gain & 0.01 \\
        $\mu$  & HPF gain& 0.1 \\
        $\tau$  & HPF time constant & 0.1 \\
        \bottomrule
    \end{tabular}
\end{table}
\subsection{Simulation Result}
{Three VSG-controlled converters are used in the simulation study, with parameters set as $J_1=D_1= 300$, $J_2=D_2=600$, and $J_3=D_3=900$. The feeder impedances are configured as \(Z_{l1} = 11{mH}\), \(Z_{l2} =7.7{mH}\), and \(Z_{l3}=6.6{mH}\).}
\subsubsection{{{Oscillation in SA mode}}}
{The simulation shows the effectiveness of the proposed method over the traditional VSG and the method in \cite{9220837,gao2024adaptive,10130744} under SA mode. The comparison study is shown in Fig.\ref{Tra_VSG}-Fig.\ref{DVI}. The active power and frequency are displayed in their subfigure (a) and (b), respectively.} 
\begin{figure*}[t]
\begin{minipage}{0.33\textwidth}
  \centering
  \subfloat[Conventional VSG under strong grid]{\includegraphics[width=1\textwidth]{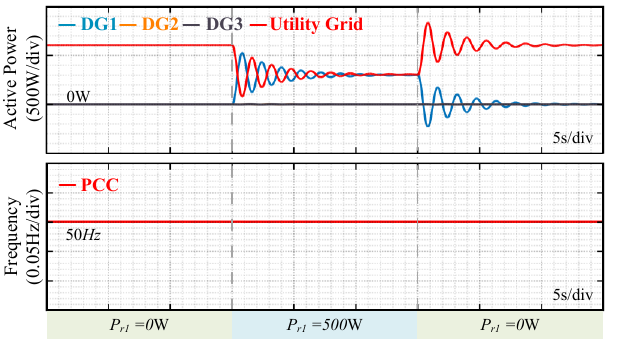}}\hfill
  \subfloat[Conventional VSG under weak grid]{\includegraphics[width=1\textwidth]{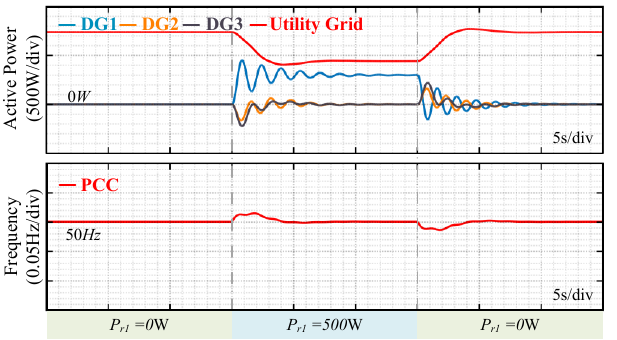}}
  \caption{{Dynamics of Conventional VSG}}
  \label{Conventional_VSG}
  \end{minipage}
  \begin{minipage}{0.33\textwidth}
  \centering
  \subfloat[Adaptive inertia under strong grid]{\includegraphics[width=1\textwidth]{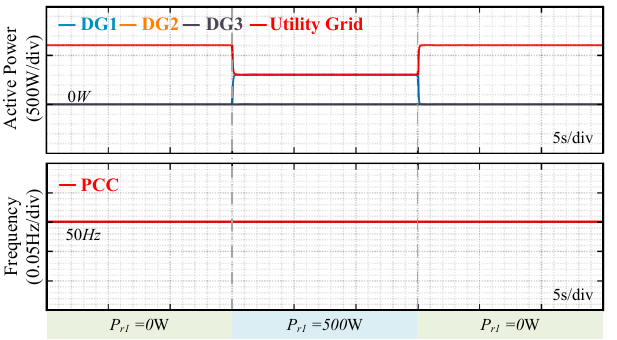}}\hfill
  \subfloat[Adaptive inertia under weak grid]{\includegraphics[width=1\textwidth]{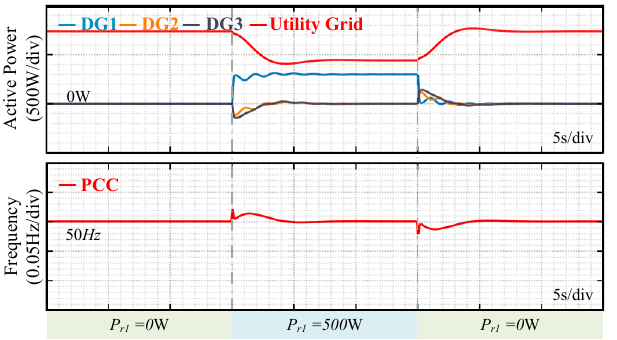}}
  \caption{{Dynamics of adaptive inertia}}
  \label{Proposed_method_A}
  \end{minipage}
  \begin{minipage}{0.33\textwidth}
  \centering
  \subfloat[Adaptive damping under strong grid]{\includegraphics[width=1\textwidth]{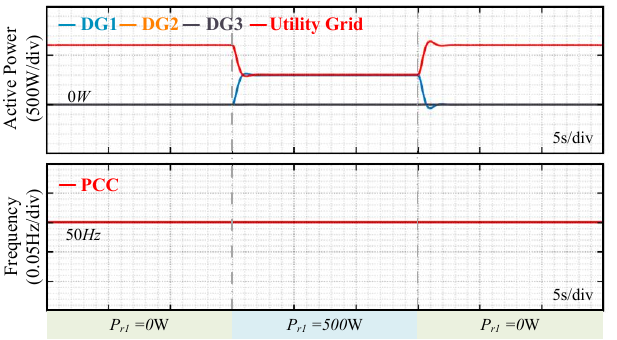}}\hfill
  \subfloat[Adaptive damping under weak grid]{\includegraphics[width=1\textwidth]{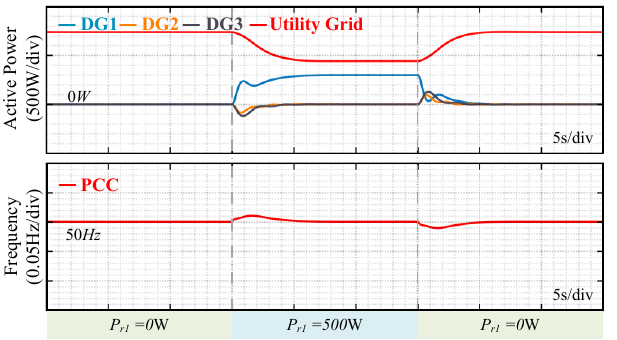}}
  \caption{{Dynamics of adaptive damping}}
  \label{Proposed_method_damping}
  \end{minipage}
\end{figure*}
\begin{figure*}[b]
\begin{minipage}{0.33\textwidth}
  \centering
  \subfloat[VSG control in SA mode]{\includegraphics[width=1\textwidth]{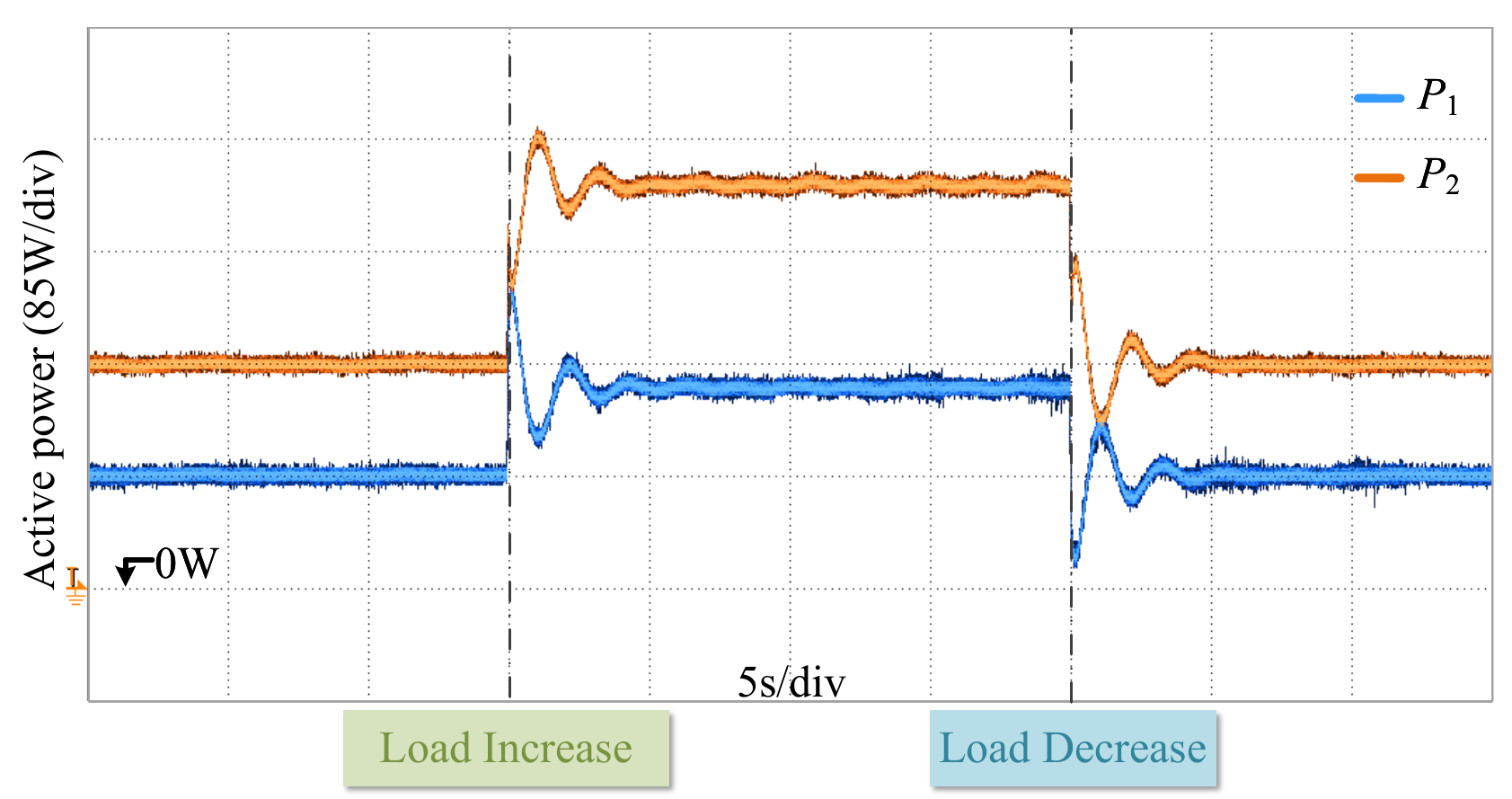}}\hfill
  \subfloat[Proposed control in SA mode]{\includegraphics[width=1\textwidth]{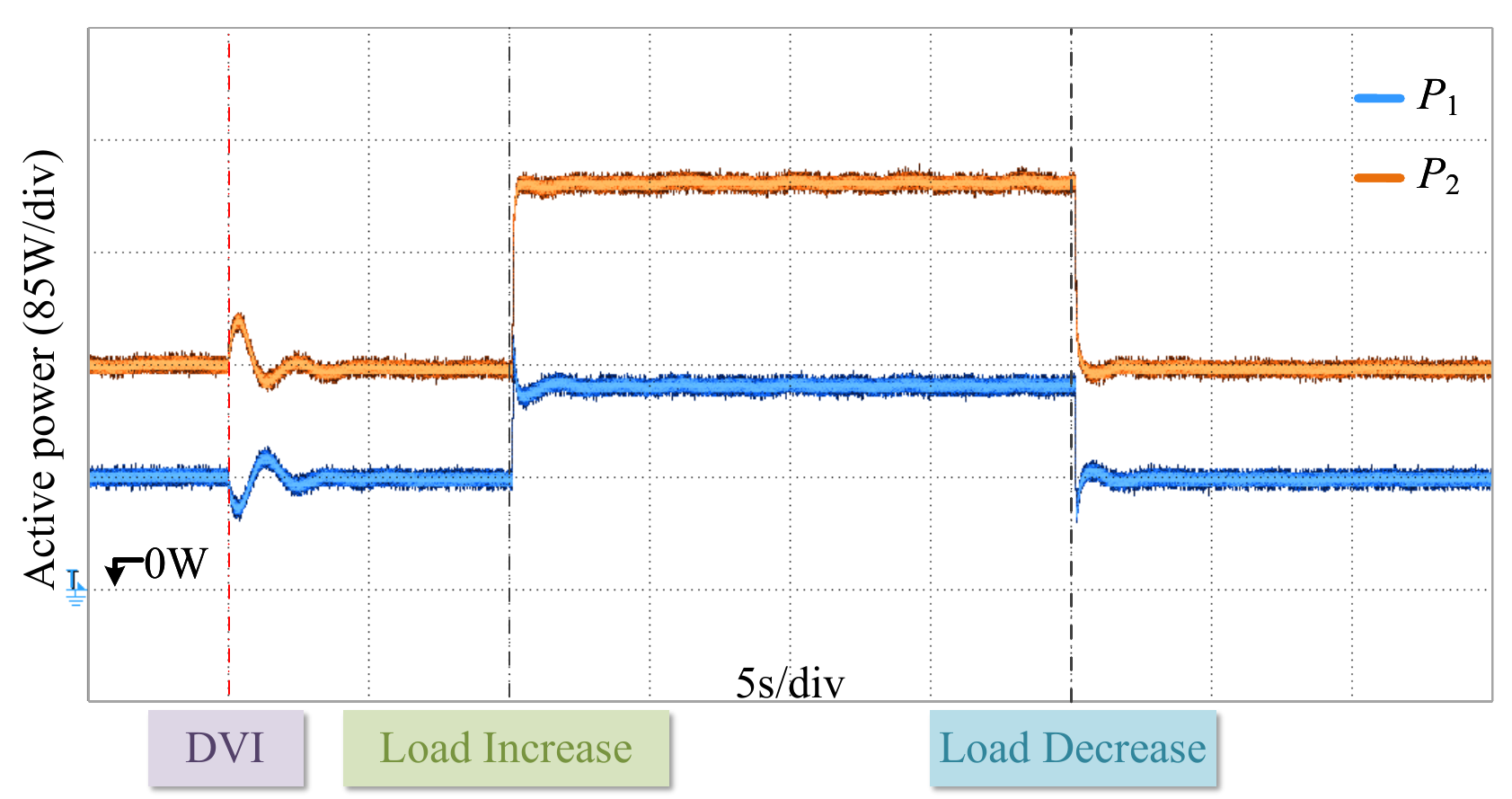}}
  \caption{{Active power comparison}}
  \label{Active_power_comparison_SA}
  \end{minipage}
  \begin{minipage}{0.33\textwidth}
  \centering
  \subfloat[VSG control in SA mode]{\includegraphics[width=1\textwidth]{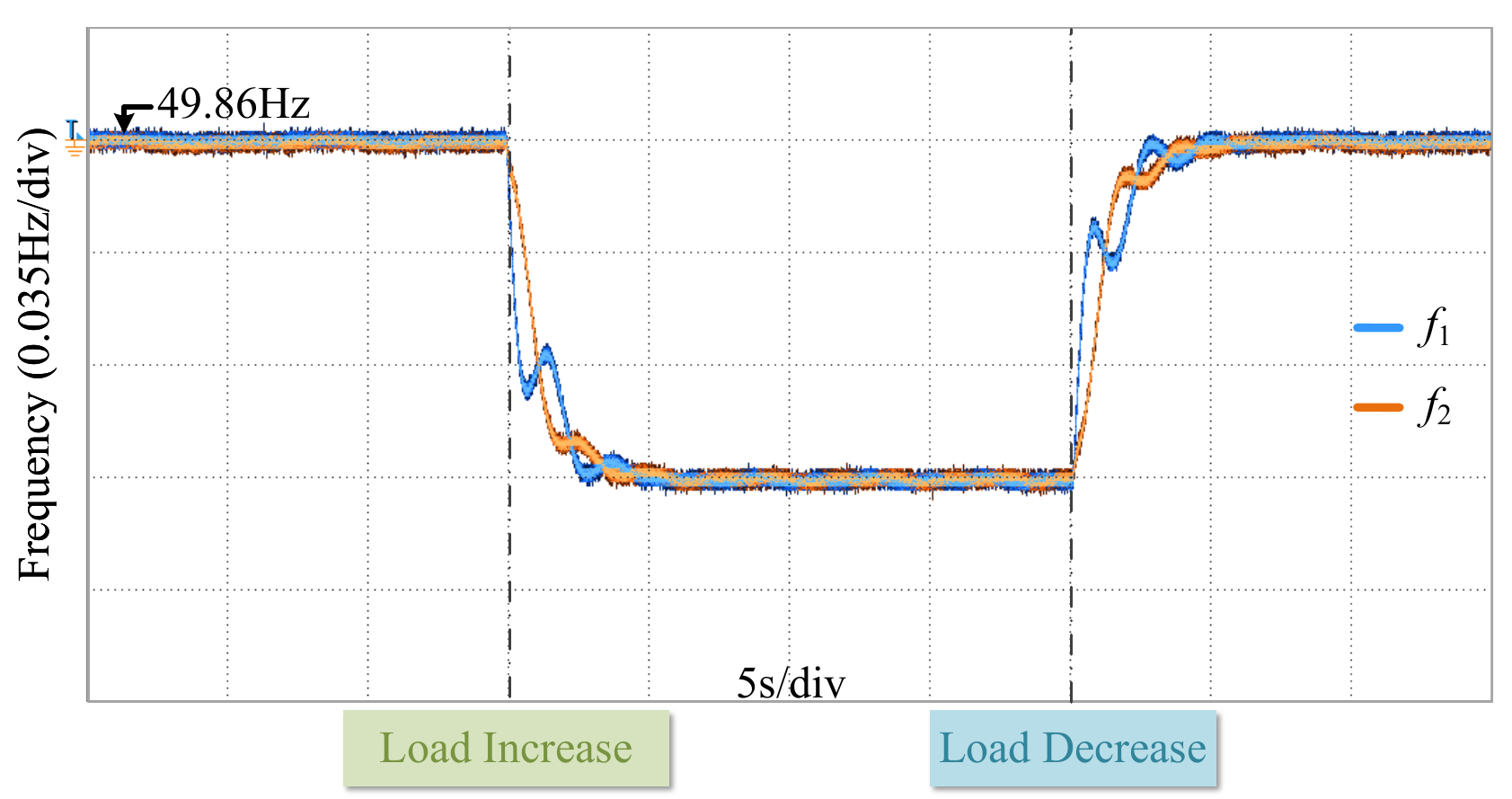}}\hfill
  \subfloat[Proposed control in SA mode]{\includegraphics[width=1\textwidth]{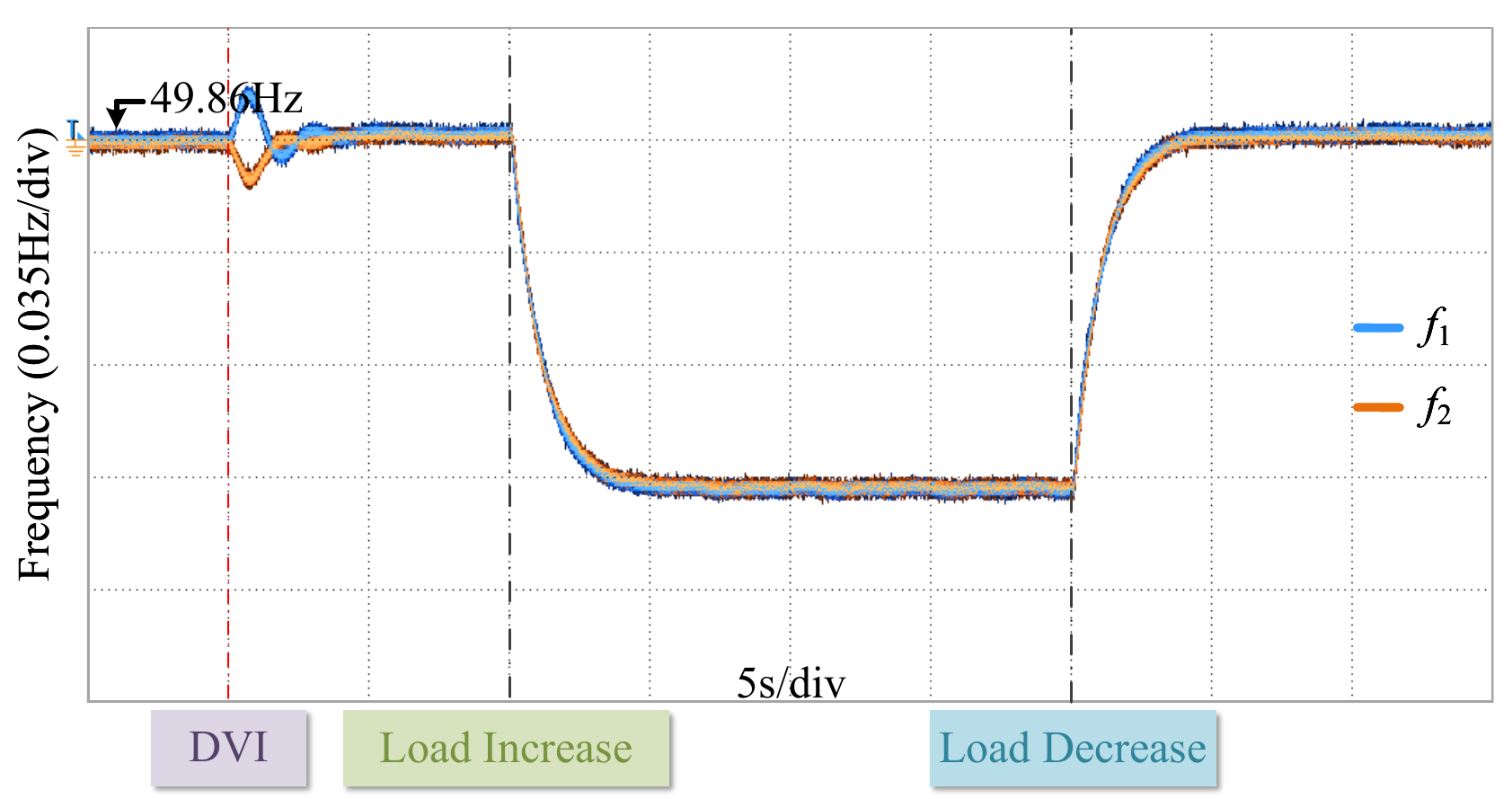}}
  \caption{{Frequency comparison}}
  \label{Frequency_comparison_SA}
  \end{minipage}
  \begin{minipage}{0.33\textwidth}
  \centering
  \subfloat[VSG control in SA mode]{\includegraphics[width=1\textwidth]{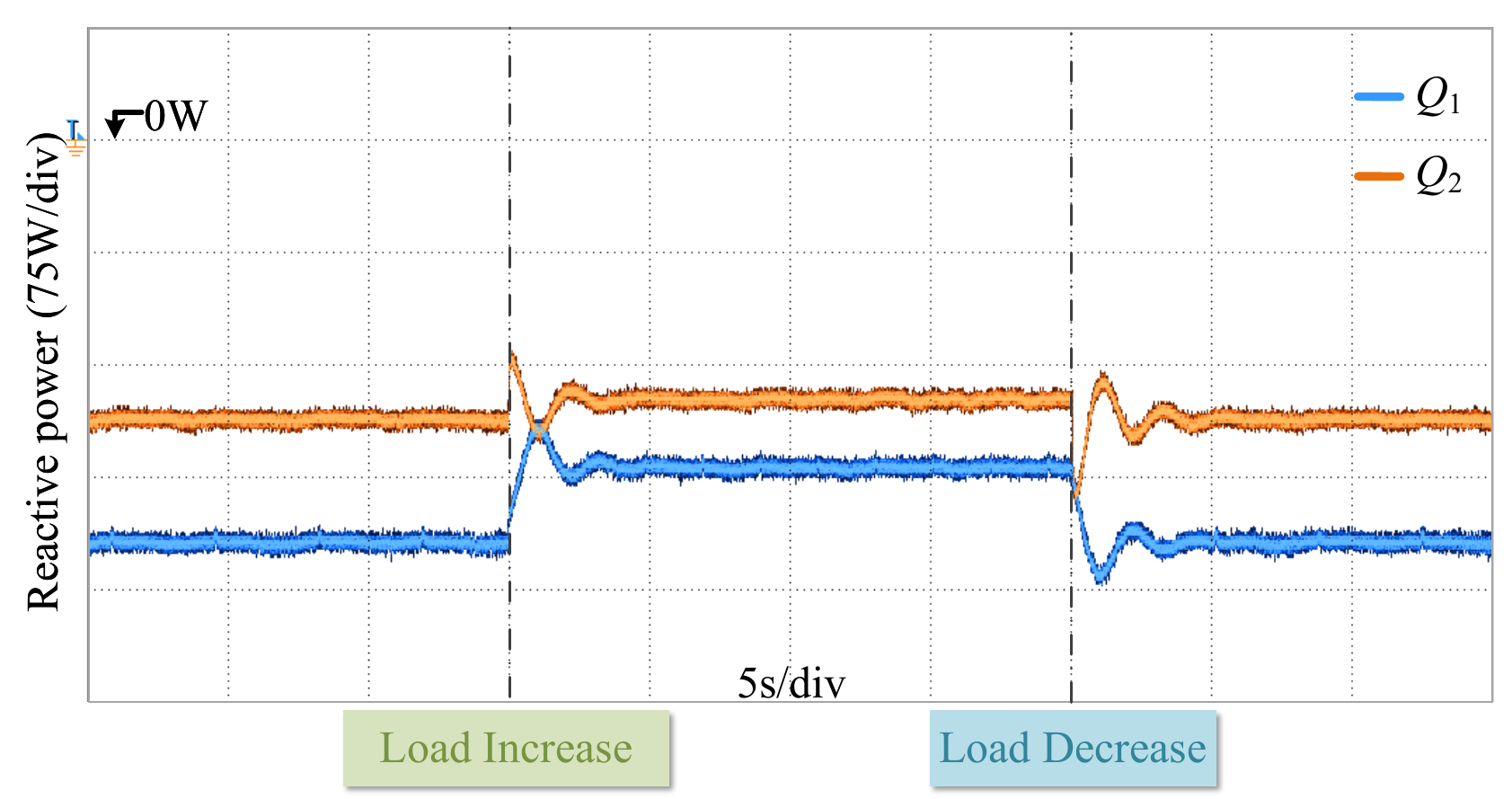}}\hfill
  \subfloat[Proposed control in SA mode]{\includegraphics[width=1\textwidth]{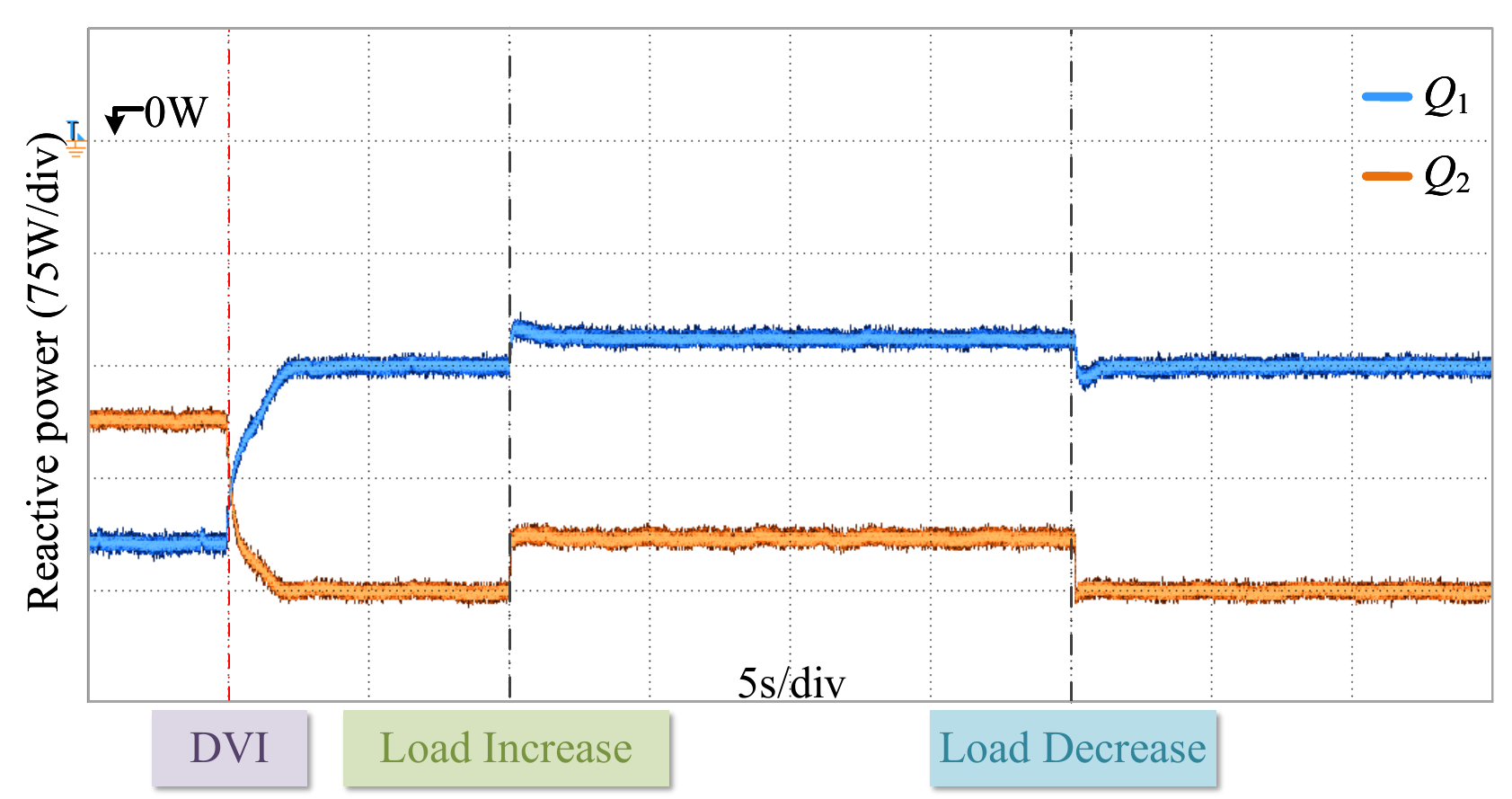}}
  \caption{{Reactive power comparison}}
  \label{Reactive_power_comparison_SA}
  \end{minipage}
\end{figure*}

{As shown in Fig.\ref{Tra_VSG}, the conventional VSG control is used to regulate the microgrids after the system starts, resulting in a proportional steady-state active power-sharing ratio of 1:2:3, as expected. Subsequently, a 700W load is added, which, according to the droop law, leads to decreased frequency. However, the active power and frequency experience severe oscillations due to the mismatched feeder impedance. When the load is suddenly switched off, the system recovers to its original power level, but oscillations persist in the dynamics.} 

{As shown in Fig.\ref{DSC}, the distributed secondary control proposed in \cite{9220837} is activated, providing extra damping for the VSG system. As seen, when the load is changed, the oscillations are relatively smaller than the conventional VSG control, as with the feedback compensation, the oscillations cannot be instantly mitigated. Since DSC necessitates a communication network, it is necessary to consider the scenario of communication loss for power converter control\cite{10979334}. Therefore,  the DSC performance is tested without communication under the load switch. Subsequently, the communication is removed, indicating that the inverters can no longer receive information from each other. In this case, when the load increases again, the active power and frequency oscillations are equivalent to those observed under traditional VSG control, indicating that DSC loses its effectiveness in mitigating oscillations. This demonstrates that DSC is not robust against communication disruptions.} {Fig.\ref{DVI} compares the proposed distributed virtual impedance and the conventional VSG control. The DVI is activated first, where a slight oscillation occurs due to the tuned impedance affecting the active power slightly. When the load is switched, the active power and frequency smoothly transition to their steady state without significant oscillations. This illustrates the effectiveness of the proposed DVI over the DSC in \cite{9220837}. Subsequently, the communication is removed for test purposes. As the DVI has been fixed and will not be changed, the parameters can remain matched for the rest of the operation. Consequently, even with load changes under the no-communication scenario, the active power and frequency do not experience oscillations. This procedure suggests that the proposed DVI control method is more immune to communication delays and interruptions than the DSC.}
\subsubsection{{{Oscillation in GC mode}}}

{We validate the proposed methods under both strong and weak grid conditions and compare three controllers: the traditional VSG, the adaptive inertia (adaptive $J$), and the feedforward-based adaptive damping (adaptive $D$). As shown in Fig.{\ref{MG}}, a {strong grid} is modeled by a stiff utility connection ($Z_{lg}=0$, equivalently $K_g\!\to\!\infty$), while a {weak grid} is emulated by adding a series grid impedance ($Z_{lg}=6Z_{l1}$). 

\textbf{Under strong grids}, the PCC frequency is fixed at 50 $Hz$, as it is clamped the utility grid. With the traditional VSG, step changes in the power reference of $unit 1$($P_{r1}$) by  500 $W$ excite noticeable active-power oscillations, see Fig.~\ref{Conventional_VSG}(a). Applying the {adaptive inertia} strategy effectively suppresses these oscillations and achieves fast and well-damped tracking while leaving the PCC frequency unchanged, as seen in Fig.~\ref{Proposed_method_A}(a). As seen in Fig.~\ref{Proposed_method_damping}(a), the adaptive damping control can mitigate oscillations but exhibits a slower transient response. \textbf{Under weak grids}, the traditional VSG yields both active-power oscillations and PCC-frequency fluctuations due to the reduced grid stiffness (Fig.~\ref{Conventional_VSG}(b)). The adaptive inertia approach attenuates power oscillations but still introduces undesired PCC frequency dynamics (Fig.~\ref{Proposed_method_A}(b)). In contrast, the adaptive damping strategy suppresses oscillations while maintaining improved PCC-frequency behavior (Fig.~\ref{Proposed_method_damping}(b)), thus offering a more balanced performance in weak-grid.

Based on the above comparisons, we recommend: (i) in \textbf{strong-grid} GC operation, use \emph{adaptive inertia} for fast and well-damped power tracking with the PCC frequency clamped by the utility; (ii) in \textbf{weak-grid} GC operation, prefer \emph{adaptive damping} to balance oscillation suppression and PCC-frequency dynamics. For completeness, we note that virtual impedance can also reduce oscillations, but overly large values incur excessive voltage drops and are therefore not advised for either strong- or weak-grid scenarios.}
\begin{figure}[h]
\centerline{\includegraphics[width=2.6in]{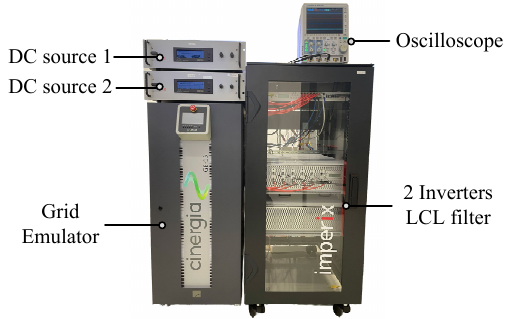}}
\caption{Experiment setup.}
\label{Experiment setup}
\end{figure}
\subsection{Experiment Result}
The proposed adaptive control strategy is further validated through experiments involving two VSG-based inverters and an ideal grid emulator. The experimental setup is illustrated in Fig.\ref{Experiment setup}. {In this setup, the control parameters are configured as $J_1=D_1=100$ and $J_2=D_2=200$. The corresponding feeder and grid impedances are \(Z_{l1}=2.2{mH}\), \(Z_{l2}=15.4{mH}\), and \(Z_{lg}=15.4{mH}\).}
\subsubsection{{Oscillation in SA mode}}
{This study's experimental results for active power, frequency, and reactive power are depicted in Fig.\ref{Active_power_comparison_SA}-Fig.\ref{Reactive_power_comparison_SA}. {The system initially undergoes a loading phase followed by an unloading phase. The load change is applied by relay-switching a resistive load at the inverter output.} The results in Fig.\ref{Active_power_comparison_SA}(a) and Fig.\ref{Frequency_comparison_SA}(a) demonstrate that significant oscillations between all sources are evident in the active power and frequency when conventional VSG control is employed. This is accompanied by disproportionate sharing of reactive power in Fig.\ref{Reactive_power_comparison_SA}(a). Notably, poor frequency dynamic performance, such as significant frequency overshoot, poses a risk of unexpected load-shedding or extensive blackouts. Fig.\ref{Active_power_comparison_SA}(b), Fig.\ref{Frequency_comparison_SA}(b), and Fig.\ref{Reactive_power_comparison_SA}(b) illustrate the effectiveness of the proposed DVI in mitigating active power oscillations. Following DVI activation, dynamic adjustments of the reactive power are observed. These adjustments result in the reactive power $Q_1$ and $Q_2$ approaching their expected values, ultimately achieving a ratio of 1:2. As discussed in the previous section, appropriate reactive power sharing implies that the equivalent impedance of the DGs has been proportionally tuned. When inertia and damping coefficients are proportionally set following the DGs' maximum output capacity and RoCoF requirements, all parameters are harmonized, thereby eliminating the oscillations caused by VSG control. This is evident in Fig.\ref{Active_power_comparison_SA}(b), where post-DVI activation, load variations do not induce active power oscillations. Furthermore, with DVI, the frequency change rate remains as expected in Fig.\ref{Frequency_comparison_SA}(b).}
\subsubsection{Oscillations in GC mode}
\begin{figure}[h]
  \centering  % This centers the entire figure
  \begin{minipage}{0.33\textwidth}
    \centering
    \subfloat[{VSG control in GC mode}]{\includegraphics[width=1\textwidth]{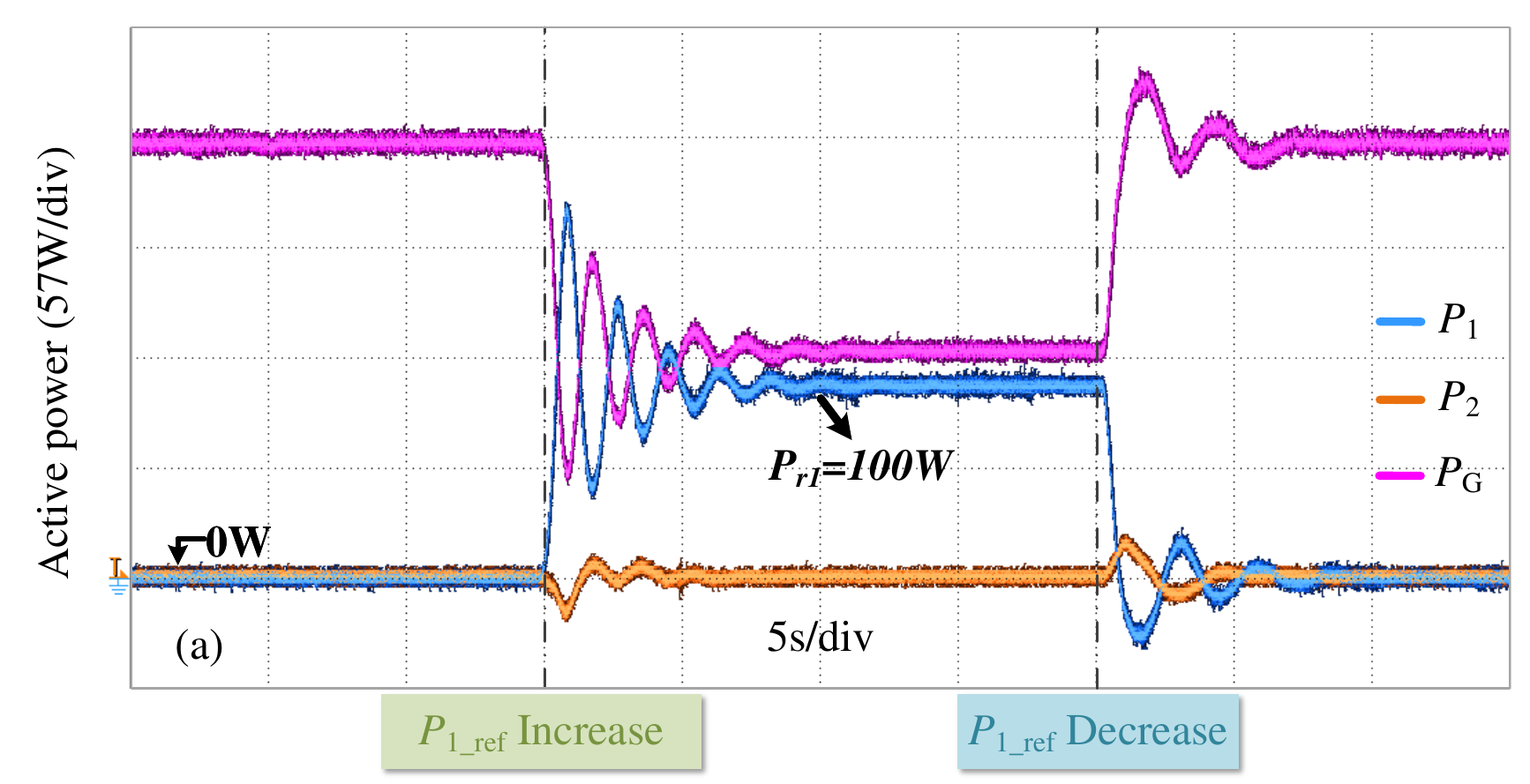}}\hfill
    \subfloat[{Proposed control in GC mode}]{\includegraphics[width=1\textwidth]{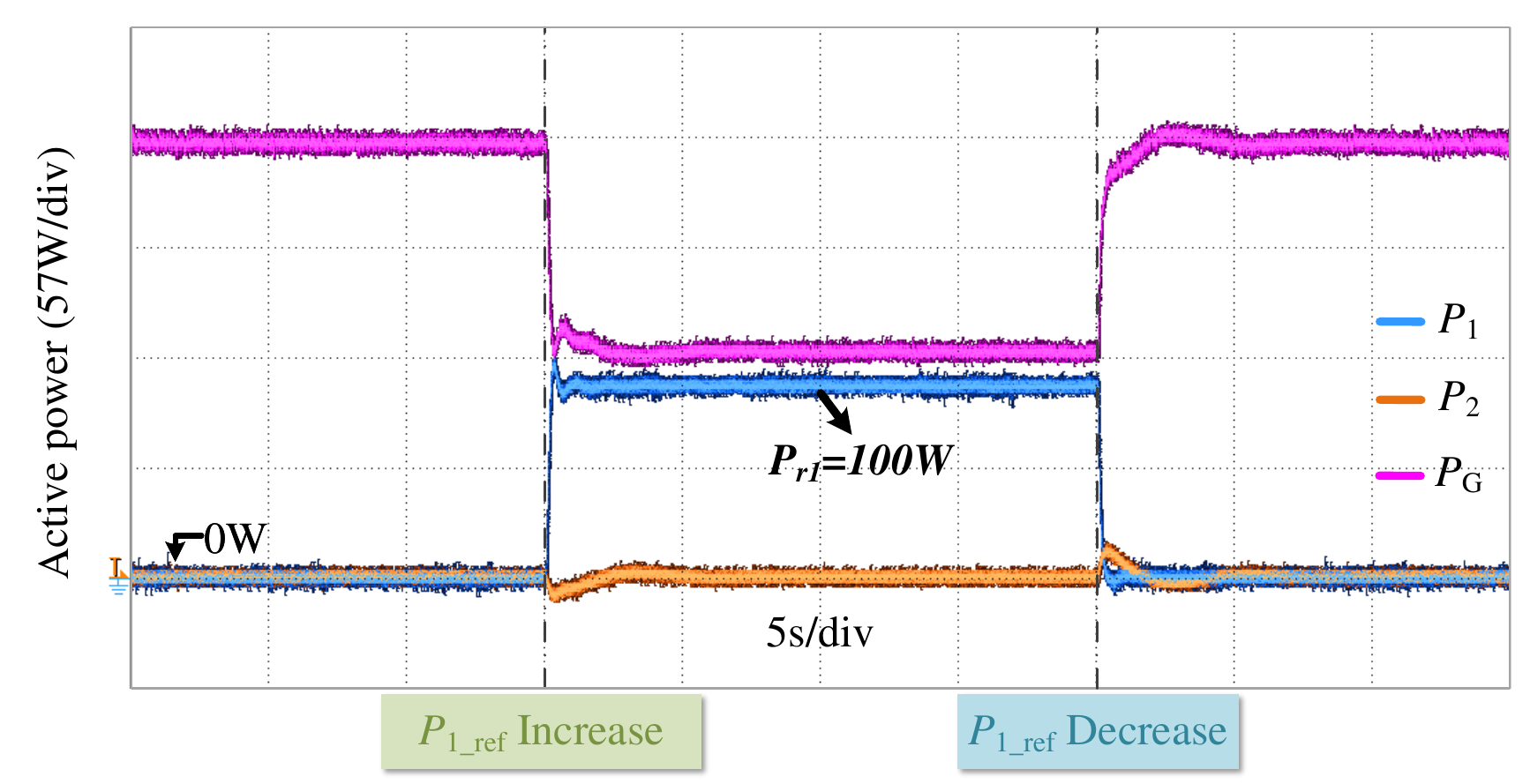}}
 \caption{{Active power comparison}}
    \label{Active_power_comparison_GC}
  \end{minipage}
\end{figure}

Initially, the DG1 and DG2 output active power is 0W, while the stiff grid supports an active power of 228W. {Consequently, a step change of 100W is applied to the power reference of the controller of DG1 in the active power set point of the grid-tied VSG to demonstrate power oscillations.} As depicted in Fig.\ref{Active_power_comparison_GC}(a), following the increase in the power set point, significant oscillations occur in the active power outputs of both DG1 and the grid. Subsequently, similar oscillations in active power are still observed when the power set point reverts to zero. Notably, the active power of DG2 remains relatively small and can be disregarded, as discussed in the previous section. Fig.\ref{Active_power_comparison_GC}(b) illustrates the dynamic performance of the proposed adaptive inertia method. It is evident that, under the same power reference change, compared to the traditional method, the adaptive inertia method enables the involved converters and the grid power to adjust smoothly to the reference without significant overshoot oscillations. The adaptive inertia is only activated by the power reference change in the GC mode, thereby not degrading the  RoCoF in the SA mode.
\section{Conclusion}
{This paper revisits VSG dynamics from an impedance–circuit perspective. In this view, inertia, damping, and feeder impedance correspond to the capacitance, resistance, and inductance of an RLC network. Active-power oscillations are then understood as LC-type resonances, excited by load or set-point disturbances. This provides an intuitive and physics-based explanation that applies to both SA and GC modes.

In SA mode,  oscillations arise from parameter mismatches across units. Moreover, this paper shows that when inertia, damping, and feeder impedance follow a proportional relationship, the oscillations can, in principle, be removed. Accordingly, we developed a distributed virtual-impedance scheme that aligns equivalent impedances across units. This suppresses oscillations without requiring feeder parameter knowledge and preserves the nominal inertia and damping property. Proportional reactive-power sharing serves as a practical indicator that the impedances are well aligned.

In GC mode, it is shown that adding inertia changes the VSG dynamics into a second-order system. This naturally introduces oscillatory transients during reference changes. To address this, we designed a feedforward-based adaptive inertia and damping for strong and weak grids, respectively. They are activated only when the power reference varies. This provides faster transient damping for power oscillation mitigation while maintaining PCC RoCoF.}
\bibliographystyle{IEEEtran}
\bibliography{ref}
\newpage
\newpage
\end{document}